\documentstyle{mn}

\newif\ifAMStwofonts

\ifoldfss
  \ifCUPmtlplainloaded \else
    \NewTextAlphabet{textbfit} {cmbxti10} {}
    \NewTextAlphabet{textbfss} {cmssbx10} {}
    \NewMathAlphabet{mathbfit} {cmbxti10} {} 
    \NewMathAlphabet{mathbfss} {cmssbx10} {} 
  \fi
  \ifAMStwofonts
    \ifCUPmtlplainloaded \else
      \NewSymbolFont{upmath} {eurm10}
      \NewSymbolFont{AMSa} {msam10}
      \NewMathSymbol{\upi}     {0}{upmath}{19}
      \NewMathSymbol{\umu}     {0}{upmath}{16}
      \NewMathSymbol{\upartial}{0}{upmath}{40}
      \NewMathSymbol{\leqslant}{3}{AMSa}{36}
      \NewMathSymbol{\geqslant}{3}{AMSa}{3E}

      \let\leq=\leqslant 
      \let\geq=\geqslant 
    \fi
  \fi
\fi 

\ifnfssone
  \newmathalphabet{\mathit}
  \addtoversion{normal}{\mathit}{cmr}{m}{it}
  \addtoversion{bold}{\mathit}{cmr}{bx}{it}
  \newmathalphabet{\mathbfit} 
  \addtoversion{normal}{\mathbfit}{cmr}{bx}{it}
  \addtoversion{bold}{\mathbfit}{cmr}{bx}{it}
  \newmathalphabet{\mathbfss} 
  \addtoversion{normal}{\mathbfss}{cmss}{bx}{n}
  \addtoversion{bold}{\mathbfss}{cmss}{bx}{n}
  \ifAMStwofonts
    \ifCUPmtlplainloaded \else
      \UseAMStwoboldmath
      \makeatletter
      \new@mathgroup\upmath@group
      \define@mathgroup\mv@normal\upmath@group{eur}{m}{n}
      \define@mathgroup\mv@bold\upmath@group{eur}{b}{n}
      \edef\UPM{\hexnumber\upmath@group}
      \new@mathgroup\amsa@group
      \define@mathgroup\mv@normal\amsa@group{msa}{m}{n}
      \define@mathgroup\mv@bold\amsa@group{msa}{m}{n}
      \edef\AMSa{\hexnumber\amsa@group}
      \makeatother
      \mathchardef\upi="0\UPM19
      \mathchardef\umu="0\UPM16
      \mathchardef\upartial="0\UPM40
      \mathchardef\leqslant="3\AMSa36
      \mathchardef\geqslant="3\AMSa3E

      \let\leq=\leqslant 
      \let\geq=\geqslant 
    \fi
  \fi
\fi 

\ifnfsstwo
  \DeclareMathAlphabet{\mathbfit}{OT1}{cmr}{bx}{it}
  \SetMathAlphabet\mathbfit{bold}{OT1}{cmr}{bx}{it}
  \DeclareMathAlphabet{\mathbfss}{OT1}{cmss}{bx}{n}
  \SetMathAlphabet\mathbfss{bold}{OT1}{cmss}{bx}{n}
  \ifAMStwofonts
    \ifCUPmtlplainloaded \else
      \DeclareSymbolFont{UPM}{U}{eur}{m}{n}
      \SetSymbolFont{UPM}{bold}{U}{eur}{b}{n}
      \DeclareSymbolFont{AMSa}{U}{msa}{m}{n}
      \DeclareMathSymbol{\upi}{0}{UPM}{"19}
      \DeclareMathSymbol{\umu}{0}{UPM}{"16}
      \DeclareMathSymbol{\upartial}{0}{UPM}{"40}
      \DeclareMathSymbol{\leqslant}{3}{AMSa}{"36}
      \DeclareMathSymbol{\geqslant}{3}{AMSa}{"3E}

      \let\leq=\leqslant 
      \let\geq=\geqslant 
    \fi
  \fi
\fi 

\ifCUPmtlplainloaded \else
  \ifAMStwofonts \else 
    \def\upi{\pi}
    \def\umu{\mu}
    \def\upartial{\partial}
  \fi
\fi

\title[Radio--galaxy evolution to $z=0.7$] {Radio galaxies in the 
2SLAQ Luminous Red Galaxy Survey: I. The evolution of low--power radio galaxies 
to $z\sim0.7$}

\author[Sadler et al. ]{
\parbox[t]{16cm}{
Elaine M.\ Sadler$^1$, 
Russell D.\ Cannon$^2$, 
Tom Mauch$^1$, 
Paul J.\ Hancock$^1$, 
David A.\ Wake$^3$,
Nic Ross$^3$,
Scott M. Croom$^1$,
Michael J.\ Drinkwater$^4$, 
Alastair C.\ Edge$^3$, 
Daniel Eisenstein$^5$,
Andrew M.\ Hopkins$^1$,
Helen M. Johnston$^1$,
Robert Nichol$^6$, 
Kevin A.\ Pimbblet$^4$, 
Roberto De Propris$^7$, 
Isaac G.\ Roseboom$^4$, 
Donald P.\ Schneider$^8$,
Tom Shanks$^3$ 
\vspace*{6pt} }\\
$^{1}$School of Physics, University of Sydney, NSW 2006, Australia \\
$^{2}$Anglo-Australian Observatory, P.O.\ Box 296, Epping, NSW 2121,
    Australia \\  
$^{3}$Department of Physics, University of Durham, South Road, Durham 
DH1 3LE \\ 
$^{4}$ Department of Physics, University of Queensland, Brisbane, QLD 4072, 
Australia\\
$^{5}$ Steward Observatory, 933 N. Cherry Ave, Tucson, AZ 85721, USA\\
$^{6}$ Institute of Cosmology and Gravitation, University of Portsmouth,
Portsmouth, PO1 2EG \\
$^{7}$ Cerro Tololo Inter-American Observatory, Casilla 63-D, La Serena, 
Chile\\
$^{8}$ Department of Astronomy and Astrophysics, Pennsylvania State University, 
525 Davey Laboratory, University Park, PA 16802, USA }
\pagerange{\pageref{firstpage}--\pageref{lastpage}}
\pubyear{2006}

\begin{document}

\maketitle

\label{firstpage}

\begin{abstract}
We have combined optical data from the 2dF-SDSS Luminous Red Galaxy and QSO (2SLAQ) 
redshift survey with radio measurements from the 1.4\,GHz VLA FIRST and NVSS surveys 
to identify a volume--limited sample of 391 radio galaxies at redshift $0.4<z<0.7$. 
By determining an accurate radio luminosity function for luminous early--type galaxies 
in this redshift range, we can investigate the cosmic evolution of the radio--galaxy 
population over a wide range in radio luminosity.  

The low--power radio galaxies in our LRG sample (those with 1.4\,GHz 
radio luminosities in the range 10$^{24}$ to 10$^{25}$\,W\,Hz$^{-1}$, corresponding 
to FR\,I radio galaxies in the local universe) undergo significant cosmic evolution 
over the redshift range $0<z<0.7$, consistent with pure luminosity evolution of 
the form (1+$z$)$^k$, where $k=2.0\pm0.3$.  Our results appear to rule out 
(at the 6--7$\sigma$\ level) models in which low--power radio galaxies undergo 
no cosmic evolution.  The most powerful radio galaxies in our sample (with radio 
luminosities above 10$^{26}$\,W\,Hz$^{-1}$) may undergo more rapid evolution over 
the same redshift range.  

The evolution seen in the low--power radio-galaxy population implies that the 
total energy input into massive early--type galaxies from AGN heating 
increases with redshift, and was at least 50\% higher at $z\sim0.55$ 
(the median redshift of the 2SLAQ LRG sample) than in the local universe. \\

\end{abstract}

\begin{keywords}
galaxies: radio continuum --- galaxies: luminosity function --- 
galaxies: active --- AGN: evolution 
\end{keywords}


\section{Introduction}

The strong cosmic evolution of the most powerful radio galaxies 
was deduced more than forty years ago from radio source counts 
(Longair 1966), which imply that the space density of powerful 
radio galaxies at redshift $z\sim2$ was roughly a thousand times 
higher than in the local universe (Doroshkevich, Longair \& Zeldovich 
1970, Dunlop \& Peacock 1990).  

Far less is known about the cosmic evolution of the 
lower--power radio galaxies which comprise the overwhelming majority 
of the local radio AGN population.  Relatively few low--power 
radio galaxies have been observed at redshifts beyond $z\sim0.3$, partly 
because classical flux--limited radio surveys like the Cambridge 3CR (Laing, 
Riley \& Longair 1983) and Molonglo MRC (Large et al.\ 1981) sample only 
a narrow range in radio luminosity at any given redshift (see e.g.\ Blundell 
et al.\ 2002); but the observed distribution of radio source counts implies 
that low--power radio sources cannot evolve as rapidly as the most powerful 
sources do (Longair 1966). 

This finding led to the development of two classes of model for the 
cosmic evolution of radio--loud active galaxies: single-population models, 
in which the rate of evolution varies 
with radio power (Dunlop \& Peacock 1990), and dual--population models, 
in which the radio--source population is assumed to be made up of a 
low--luminosity non--evolving component and a high--luminosity 
rapidly--evolving component (Jackson \& Wall 1999). In the 
dual--population models, the two populations have been variously been 
identified 
with FR\,I and FR\,II radio galaxies\footnote{Fanaroff \& Riley (1974) 
divided radio galaxies 
into two classes based on the observed morphology of their radio emission 
and showed that this morphology correlates with radio luminosity, with 
the less luminous (FR\,I) objects having a jet-like appearance and the 
more luminous (FR\,II) objects having edge-brightened radio hotspots.  
Later work 
(Bicknell 1995) showed that the classification appears to have a physical 
basis, since FR\,II radio galaxies have jets which remain relativistic over 
scales of tens to hundreds of kiloparsec, whereas FR\,I radio jets rapidly 
decelerate to sub-relativistic velocities. As a result, models of radio-galaxy 
evolution often treat FR\,I and FR\,II radio galaxies as separate populations 
which may evolve in different ways.  } (Wall 1980; later expanded by Jackson \& Wall 
1999 to include BL Lac objects and flat--spectrum QSOs as the beamed 
counterparts of FR\,I and FR\,II objects respectively); or with 
objects having weak and strong optical emission lines, independent 
of their radio morphology (Willott et al.\ 2001). 

Several recent studies imply that low--power radio galaxies 
(i.e.\ those with 1.4\,GHz radio luminosities near or below the 
FR\,I/FR\,II divide at $\sim10^{26}$\,W\,Hz$^{-1}$) undergo little or no 
cosmic evolution.  In the model of Jackson \& Wall (1999), which is consistent 
with both radio source counts and the observed redshift distribution of 3CR 
sources, there is strong cosmic evolution of FR\,II radio galaxies 
but no evolution of the FR\,I population.  Clewley \& Jarvis (2004) 
also found no increase in the comoving density of radio sources with 
luminosities below P$_{\rm 325\,MHz}\sim10^{25}$\,W\,Hz$^{-1}$ over 
the redshift range $0<z<0.8$.  

In contrast, Brown, Webster \& Boyle (2001) found significant 
luminosity evolution (of the form $(1+z)^k$, where $3<k<5$) 
for a sample of low--power radio galaxies over the redshift range 
$0<z<0.5$.  Snellen \& Best (2001) found two distant FR\,I 
radio galaxies in the small area of sky covered by the Hubble Deep Field; 
from which they argue that FR\,I radio galaxies must be significantly 
more abundant at $z>1$ than in the nearby universe.  Willott et al.\ (2001) 
also suggest that the comoving space density of low--luminosity radio 
galaxies rises by about 1 dex between $z\sim0$ and $\sim1$. 

The tight relation between the mass of luminous early--type\footnote{Throughout 
this paper, we use the term ``early--type galaxy'' to refer to both luminous, 
passively--evolving distant galaxies (LRGs) and giant E/S0 galaxies in the local 
universe. 
} 
galaxies and the mass of their central supermassive black holes (Magorrian et 
al.\ 1998) implies that the evolution of galaxies and their central black 
holes are intimately related.  There is also increasing evidence that 
radio jets can regulate and prevent star formation in luminous early--type 
galaxies by heating the interstellar gas and stopping the onset of 
cooling flows (Binney \& Tabor 1995; Rawlings \& Jarvis 2004; Birzan et 
al. 2004; Springel, Di Matteo \& Hernquist 2005).  Since most radio galaxies 
have radio luminosities well below the FR\,I/FR\,II break for most of 
their lifetimes, improving our understanding of the cosmic evolution 
of these lower--power radio galaxies is an essential first step 
in understanding their effects on the star--formation history of 
massive galaxies.  This is particularly important in the context of 
recent semi--analytic models of galaxy evolution which incorporate 
AGN heating (e.g. Croton et al. 2006; Bower et al. 2006), as we discuss
in \S6.4 of this paper. 

Previous studies in this area have relied strongly on the analysis of 
radio source--counts and/or on the use of photometric redshifts to 
derive distances. However, the most direct and accurate way to measure 
the cosmic evolution of low--power radio galaxies is to compare the 
radio luminosity function (RLF) observed at different redshifts.  
This requires a large volume--limited (rather than flux--limited) 
sample of radio sources which is reasonably complete over a wide 
range in radio power.
Such samples can now be assembled by combining 
a large optical redshift survey with a sensitive, large--area 
radio continuum survey; this technique has been used to 
measure accurate RLFs for galaxies in the local universe  
(Sadler et al.\ 2002; Best et al.\ 2005a,b; Mauch \& Sadler 2006).  
In this paper, we extend 
the same technique to higher redshift ($0.4<z<0.8$) by using optical 
spectra from the 2SLAQ LRG survey (Cannon et al.\ 2006) and radio data 
from the VLA FIRST (Becker et al.\ 1995) and NVSS (Condon et al.\ 1998) 
surveys.  

Throughout this paper, we use $H_0$ = 70 km s$^{-1}$ 
Mpc$^{-1}$, $\Omega_m$ = 0.3 and  $\Omega_\Lambda$ = 0.7.

\section{The 2SLAQ LRG survey}

During 2003--5, the 2SLAQ (2dF--SDSS LRG And QSO) survey 
used the 2dF multi--object spectrograph on the Anglo--Australian Telescope 
to obtain optical spectra of over 11,000 luminous red galaxies (LRGs) with 
$i$-band magnitude $<$19.8 and redshifts in the range $0.4<z<0.8$.  
Full details of the source selection, sample properties and spectroscopic 
observations are discussed by Cannon et al.\ (2006), and only a brief 
outline is given here. 

\begin{figure}
\vspace*{8cm}
\includegraphics{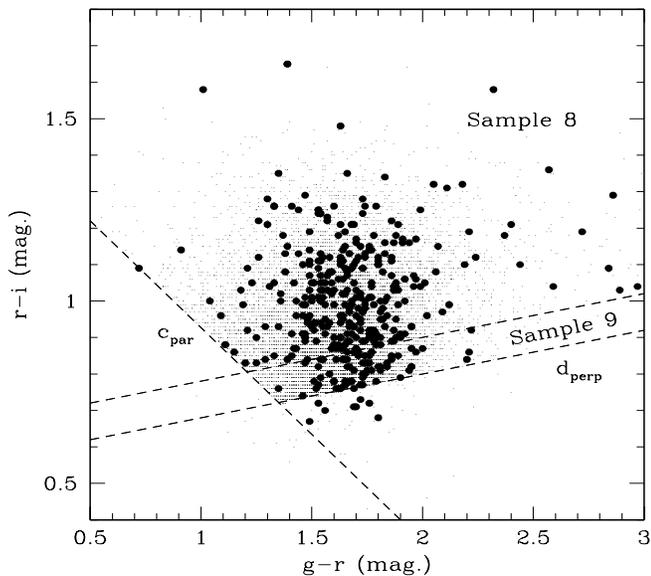}
\caption{Photometric selection criteria for the 2SLAQ LRG sample.  
Small dots show the full $\sim$15,000-galaxy spectroscopic sample 
and large filled circles the $z>0.4$ radio--detected LRGs  
discussed in \S3. The regions occupied by Samples 8 and 9 (as defined by Cannon 
et al.\ 2006) are also labelled, along with the colour--cut lines $c_{\rm par}$ and 
$d_{\rm perp}$ used to select them. A small number of 2SLAQ galaxies belong to 
other photometric samples, and fall outside the area covered by Samples 8 and 9. }
\label{fig.phot}
\end{figure} 

The total co-moving volume sampled by the 2SLAQ LRG survey out to 
the median redshift of $z$=0.55 is just over 10$^8$\,Mpc$^3$, 
i.e. a larger volume than that sampled by the 2dF Galaxy Redshift 
Survey (2dFGRS; Colless et al.\ 2001), which covered a larger area 
of sky to a shallower redshift limit (up to $z\sim0.3$ for the most 
luminous galaxies).  

\subsection{Colour selection of 2SLAQ LRGs }
The 2SLAQ LRG sample was selected using $ugriz$ (Fukugita et al.\ 1996) 
photometry from the Sloan Digital Sky Survey (SDSS; York et al.\ 2000) 
in two narrow strips along the celestial equator covering a total area 
of about 180\,deg$^2$.  
The colours used for selection were the SDSS extinction-corrected 
modelMag colours as described by Stoughton et al.\ (2002), and 
the colour selection criteria were similar 
to those used by Eisenstein et al.\ (2001) to select SDSS LRGs, 
modified as described by Cannon et al.\ (2006) to select 
target objects in the redshift range $0.4<z<0.8$. 
A (dereddened) cutoff magnitude of $i$=19.8 was applied along 
with the cuts in $g-r$ and $r-i$ colour. 

Figure \ref{fig.phot} shows the colour cuts used to select 
target 2SLAQ LRGs for the spectroscopic survey.  As discussed 
in detail by Cannon et al.\ (2006), $0.4<z<0.7$ LRGs whose 
light is dominated by an old, passively--evolving stellar population 
are expected to lie along a vertical track in this diagram, 
with $g-r\simeq1.7$.  In this redshift range, the $r-i$ colour 
of early--type galaxies becomes rapidly redder with increasing 
$z$ as the 4000$\AA$ break moves through the band, whereas 
the $g-r$ colour remains almost constant. 

The dashed line sloping downward from left to right in 
Figure \ref{fig.phot}, defined by setting a constant value 
of 1.6 for the quantity $$c_{\rm par}=0.7(g-r)+1.2(r-i-0.18),$$  
is used to separate LRGs from star--forming galaxies, which 
lie to the left of this line at $0.4<z<0.7$. 
The lines sloping upwards from left to right in Figure \ref{fig.phot}, 
defined by constant values of $$d_{\rm perp}=(r-i)-[(g-r)/8.0]$$
select early--type galaxies at increasingly high redshift for larger 
values of $d_{\rm perp}$ (Eisenstein et al.\ 2001; Cannon et al. 2006)

Two main samples of LRGs were observed.  The primary sample 
(Sample 8) consists of objects with $z>0.45$ at a surface density 
such that most targets can be accessed in a single pass.  This 
sample has over 90\% completeness both spatially and in terms of 
redshift reliability.  A secondary sample (Sample 9) contains mostly 
lower-redshift objects with $z\sim0.45$; it is photometrically 
homogeneous and has high redshift completeness but very variable 
spatial coverage. 

The photometric selection technique worked very successfully; 
over 90\% of the objects observed were LRGs in the target redshift 
range.  As discussed by Wake et al.\ (2006), these galaxies lie 
well above the `knee' in the optical luminosity function and  
have $r$--band luminosities in the range 2--15\,L$^*$.  

\subsection{2dF spectroscopy }
\begin{figure*}
\centering
\begin{minipage}{175mm}
\vspace*{11cm}
\includegraphics{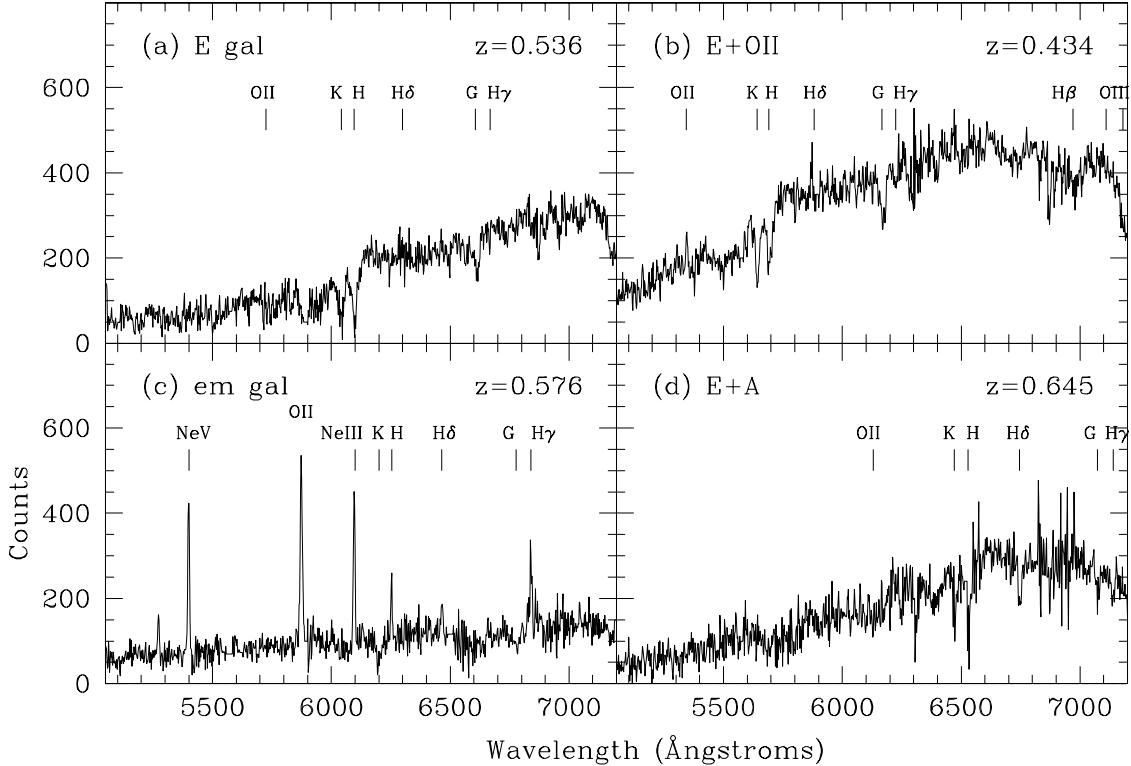}
\caption{Sample 2dF spectra of some 2SLAQ radio galaxies from Table 3. 
J144541.59$-$003409.6 at $z=0.5360$ (a) is an absorption--line galaxy 
at about the median redshift of the survey and is typical of the majority of 
2SLAQ radio galaxies, which show no obvious optical emission lines. 
J143000.12$+$001555.2 at $z=0.4343$ (b) is a lower--redshift galaxy 
with a weak [O\,II] emission line.  
J225439.77$-$001501.2 (c) is an example of an emission--line galaxy with strong 
lines of [O\,II], [Ne\,III] and [Ne\,V] and weaker Balmer emission lines of H$\gamma$ 
and H$\delta$, while  J093142.54$-$000306.1 (d) shows the strong H$\delta$ 
absorption feature characteristic of post--starburst `e+A' galaxies. 
The effective spectral resolution is about 5\,\AA. }
\label{fig.spectra} 
\end{minipage}
\end{figure*}

All the 2SLAQ spectra were obtained with the 2dF fibre spectrograph 
(Lewis et al.\ 2002), and the observing and reduction techniques 
are described in detail by Cannon et al.\ (2006).  
The wavelength coverage of the 2dF spectra was 
typically 5000--7250\,\AA, as shown in Figure \ref{fig.spectra}. 
In the redshift range targeted by the 2SLAQ survey, the main 
features seen in the 2dF spectra are the Ca\,II H and K absorption lines 
and 4000\AA\ break.  

For 2SLAQ galaxies with redshifts above $z\sim0.45$, the H$\beta$ line is 
shifted out of the 2dF wavelength range and the only observable strong 
emission line is [O\,II] 3727\AA, which is seen in the spectra of 
just over 25\% of the radio--detected 2SLAQ LRGs. 
Higher--excitation emission lines of [Ne\,V] 3426\AA\ and [Ne\,III] 3870\AA\ 
are also detected in a few 2SLAQ LRGs (like J092203.20$-$004443.5 in 
Figure \ref{fig.spectra}), as discussed later in \S4.3.  
The great majority ($>70\%$) of radio--detected 
2SLAQ LRGs show absorption--line spectra like the galaxy in Figure \ref{fig.spectra}(a), 
and only a handful have an e$+$A spectrum like that shown in \ref{fig.spectra}(d).

\section{Radio source identifications} 
We adopted a two--step approach similar to that described by Best et al.\ (2005a) 
to identify radio sources associated with the 2SLAQ LRGs, using radio--source catalogues 
from both the VLA FIRST survey (Becker et al.\ 1995) and the NVSS (Condon et al.\ 1998)  

The FIRST and NVSS radio surveys have complementary properties.  NVSS, with a 
45\,arcsec beam, accurately samples the total flux density of extended radio
sources.  FIRST, with a 5\,arcsec beam, has higher spatial resolution but 
at the expense of resolving out extended radio emission on scales larger than 
a few arcseconds (and hence underestimating the total flux density of 
galaxies with extended radio components).  We therefore use the FIRST positions 
to identify 2SLAQ radio galaxies and the NVSS flux densities to calculate the 
radio luminosity function. 

\subsection{Background}
The surface density of bright galaxies (B$<$19.5\,mag.) is low enough that reliable radio 
identifications can usually be made from the NVSS survey alone (see e.g. Sadler et al.\ 2002, 
Mauch \& Sadler 2006).  For fainter objects, however, more accurate radio positions are needed. 
Best et al.\ (2005a) have developed a multi-stage method using information from both FIRST and 
NVSS to produce a sample of radio--source IDs with high completeness and reliability, and have 
used this to identify the radio counterparts of galaxies ($14.5<r<17.8$\,mag.)
in the second data release of the SDSS.  

Experience shows that the measured radio luminosity function of volume--limited 
galaxy samples such as the SDSS and 2dFGRS is very robust against small changes in the 
way the sample is selected.  For example, the RLFs measured by Best et al.\ (2005a) 
for local AGN and star--forming galaxies agree well with those measured 
by Sadler et al.\ (2002) and Mauch \& Sadler (2006) even though slightly different 
radio identification criteria were used in all three investigations.  

In this study we use a similar approach to that of Best et al.\ (2005a), though with some 
modifications as described below.   In particular, we check all our candidate FIRST radio IDs 
visually on radio--optical overlay plots.  This matches the procedure used by Mauch \& Sadler 
(2006) to identify a large sample of radio galaxies from the 6dF Galaxy Survey 
(Jones et al.\ 2004), which we use here as the local benchmark to measure the redshift 
evolution of the radio luminosity function. 

We carried out the radio--source identification in three stages: 
\begin{itemize}
\item
Early in the project, we performed a series of tests on the 2003 version 
of the 2SLAQ LRG input photometric catalogue.  This allowed us  
to estimate both the reliability of the final radio sample and the expected number 
of extended double and multiple radio sources, as described in \S3.2.1. 
\item
Once the 2SLAQ spectroscopic observations were complete, we cross-matched the 
final spectroscopic catalogue with the FIRST survey.  
This allowed us to identify a sample of 367 2SLAQ LRGs associated with 
FIRST radio sources (see \S3.3.1). 
\item 
We also cross-matched the 2SLAQ LRG spectroscopic catalogue with the NVSS catalogue 
as discussed in \S3.3.2.  This allowed us to identify six very extended radio sources 
with angular sizes greater than 1\,arcmin, as well as 22 weak radio sources  
which were not listed  in the FIRST catalogue.  The NVSS catalogue provides accurate 
total flux densities for most of the sources identified in the FIRST catalogue.
\end{itemize} 

\subsection{Tests using the 2SLAQ LRG photometric catalogue }
\subsubsection{Radio--source identification} 
Of the 70,582 galaxies listed in the 2003 2SLAQ LRG input photometric catalogue, 
60,290 (85\%) lay within the region of sky covered by the FIRST radio catalogue.  
We checked for possible matches of these galaxies with FIRST 
radio sources, using a 30\,arcsec maximum offset between the radio and optical 
positions (because both our own tests and the study by Best et al.\ (2005a) showed 
that only a handful of genuine radio IDs are expected at larger separations).  
For the redshift range covered by the 2SLAQ LRGs ($0.4<z<0.8$), 30\,arcsec 
corresponds to a projected linear distance from the optical galaxy of 
160--220\,kpc.  The few radio galaxies with radio lobes more distant than this 
are best identified in the lower-resolution NVSS images as discussed below.  

\begin{figure}
%
%
\vspace*{8cm}
\includegraphics{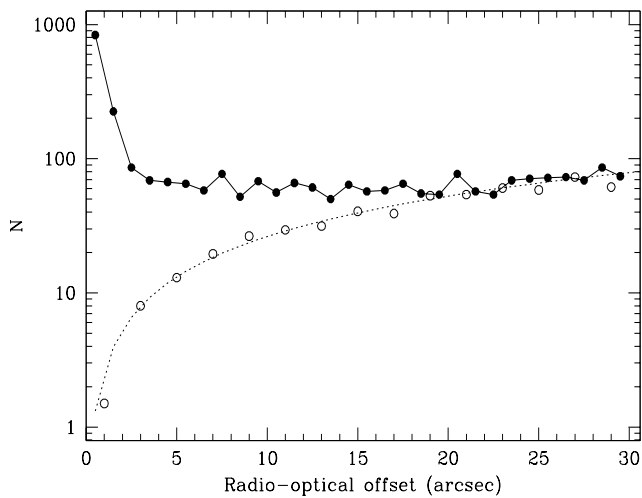}
\caption{Number of candidate FIRST radio detections of 2SLAQ LRGs (in successive 
1-arcsec annuli) plotted
against the offset $\Delta$ between the radio and optical positions. 
Filled circles represent candidate radio IDs from the 2SLAQ photometric catalogue.  
Open circles show matches found to positions randomly offset from the 2SLAQ objects, 
and allow us to estimate the optimal search radius for radio IDs.  The dashed line 
shows the number of chance coincidences expected if the FIRST radio sources are 
uniformly distributed on the sky with a surface density of 90\,deg$^{-2}$. }
\label{fig.offset}
\end{figure} 

A total of 2,782 of the LRGs in the input catalogue (4\%) had one or more FIRST radio 
sources within 30\,arcsec of the optical position, and so were candidate 
radio galaxies.  We repeated the matching 
process using a set of `random' positions offset by 10\,arcmin from the position 
of each galaxy in the 2SLAQ input catalogue. This Monte Carlo test allows 
estimation of the number of unrelated foreground or background radio sources which are 
seen by chance.  Figure \ref{fig.offset} plots the number of sources seen in 2SLAQ and 
random fields as a function of the offset between radio and optical positions.  
Note that the vertical axis is logarithmic, so the excess of sources at separations 
$\leq3$\,arcsec is very large.  As expected, the number of matched sources 
approaches the value expected by chance for offsets larger than about 20\,arcsec. 

The population of FIRST radio sources within 30\,arcsec of a 2SLAQ LRG will be  
a mixture of the following: \\
(a) Single radio sources which are genuinely associated with a 2SLAQ 
galaxy.  \\
(b) Components of double or triple radio sources which are genuinely associated with a 2SLAQ galaxy. \\
(c) Unrelated foreground or background radio sources.  The likely numbers of such objects 
can be estimated from Monte Carlo tests. \\
(d) Components of single, double or triple radio sources associated 
with neighbouring galaxies at the same redshift as the 2SLAQ galaxy.  

For each 2SLAQ LRG with a candidate radio source within 30\,arcsec, 
we overlaid radio contours from the FIRST survey onto greyscale optical 
images (taken mainly from 
the SDSS DR3, with a small number from the SuperCOSMOS images). These overlays were 
then inspected by at least two team members, who flagged each candidate source as 
`accept' or `reject' based on the following guidelines:  \\
(i) All sources less than 3.0\,arcsec from a 2SLAQ galaxy were accepted as genuine IDs. \\
(ii) A single FIRST source separated by 3--10\,arcsec from a 2SLAQ galaxy is 
accepted as an ID if (a) it is spatially resolved and extended in the direction of the 
optical galaxy, (b) the separation is no larger than the projected major axis of the radio 
source, and 
(c) no other optical object closer to the radio position is visible on the overlay images. \\
(iii) Two FIRST components of similar flux density are accepted as IDs if the 
optical galaxy lies within 5.0\,arcsec of the radio centroid (unless another optical object 
is closer). \\
(iv) Where three or more FIRST components are present, a decision on whether each is associated 
with the optical galaxy is based on visual inspection alone. 

\subsubsection{The effects of clustering} 
Of the 2,871 FIRST matches with the 2SLAQ input photometric catalogue,  1602 (56\%) were 
classified as genuine associations with 1362 2SLAQ galaxies.  This corresponds to a radio 
detection rate of 2.3\% for the 2003 2SLAQ input catalogue. Of the accepted radio galaxies, 
87\% had a single FIRST component, 10\% were doubles and 3\% were resolved into three or more 
FIRST components.  
Our Monte Carlo tests imply that the excess of `real' over `random' sources in this sample 
should be roughly 1750$\pm$40.  This is significantly higher than the 1602 sources we 
accepted as genuine associations with 2SLAQ galaxies, and at first glance might suggest that 
we have failed to recognize up to 150 genuine matches of FIRST radio sources with galaxies in 
the 2SLAQ input catalogue.  

However, as noted by Best et al.\ (2005a), Monte Carlo tests will not give a reliable estimate 
of completeness if the 2SLAQ galaxies are strongly clustered.  If the overall space 
density of galaxies is higher in the vicinity of a 2SLAQ LRG, then the probability of 
finding a radio source within 30\,arcsec of the LRG will also be higher.  For the 2003 
2SLAQ input catalogue, we have a statistical excess of 148$\pm$40 radio sources which are not 
identified with 2SLAQ galaxies but also cannot be explained by chance associations of foreground 
or background objects.  If these excess radio sources are associated with LRGs which are clustered 
near 2SLAQ galaxies but fall below the $i<19.8$\,magnitude cutoff of the 2SLAQ catalogue, 
and if we assume the same 2.3\% radio detection rate for these slightly fainter LRGs as for the 
2SLAQ galaxies, we would require an excess of about 6500 LRGs within 30\,arcsec of a 2SLAQ 
LRG. This in turn implies that $\sim$11\% of 2SLAQ LRGs have another LRG  within 
30\,arcsec on the sky (a projected separation of 190\,kpc for redshift $z$=0.55), 
which is consistent with the observed level of clustering in the 2SLAQ LRG sample 
(Ross et al.\ 2006). 


\subsection{Identification of radio galaxies in the 2SLAQ spectroscopic catalogue}

\subsubsection{Cross-matches with the FIRST and NVSS radio--source catalogues} 
The 2SLAQ LRG input catalogue was revised and expanded in 2004, so the final 
LRG spectroscopic sample is not a simple subset of the 2003 input catalogue 
discussed above.  
We therefore repeated the matching process described in \S3.2.1 for the 
14,978 galaxies in the final spectroscopic LRG sample.  The overlap with 
the FIRST survey area was significantly higher for the final spectroscopic 
catalogue (96.5\%) than for the 2003 input catalogue (85\%), giving us 
near--complete radio data for the spectroscopic sample\footnote{The fraction 
of galaxies with Galactic latitude $|b|<20^\circ$ is significantly lower in the final 
spectroscopic catalogue than in the 2003 input catalogue.  Since the FIRST survey  
has incomplete coverage of regions within 20 degrees of the Galactic Plane, its overlap 
with the final spectroscopic catalogue is correspondingly higher than the 85\% overlap 
with the 2003 input photometric catalogue.}. 

The lower--resolution NVSS is more sensitive to large--scale radio emission than the 
FIRST survey, and so provides a better measure of the total flux density of extended 
radio sources (Condon et al.\ 1998).  We independently cross-matched the 
2SLAQ spectroscopic catalogue with the NVSS, both to measure more accurate total 
flux densities for sources identified through matching with FIRST 
and also to search for extended or low surface--brightness radio sources 
which might have been missed in the FIRST survey.

Following Best et al.\ (2005a), we searched for NVSS radio components within 3\,arcmin 
of the optical position of each LRG. This search radius is large enough to include both 
lobes of any extended or multi-component radio galaxy, but smaller than the 10\,arcmin 
separation typical of unrelated NVSS sources.  It is important to note, however, that 
the probability of finding an unrelated NVSS source within 3\,arcmin of an arbitrary 
position on the sky is high ($>$30\%, see Fig.\ 6 of Condon et al.\ 1998), and even 
the probability of finding two unrelated sources within 3\,arcmin is $\sim$10\%. 

Of the 14,978 galaxies in the final LRG spectroscopic catalogue, 4378 had a single NVSS 
source within 3\,arcmin of the optical position, and a further 1316 had two or more NVSS 
sources within 3\,arcmin.  We excluded all the single-component matches with offsets 
larger than 30\,arcsec, since Mauch \& Sadler (2006) 
found that genuine matches of nearby galaxies with a single NVSS source all have 
radio--optical separations smaller than 30\,arcsec.  This left 375 candidate 
single-component NVSS matches in addition to the 1316 candidate multi-component matches. 

\begin{figure}
\vspace*{8cm}
\includegraphics{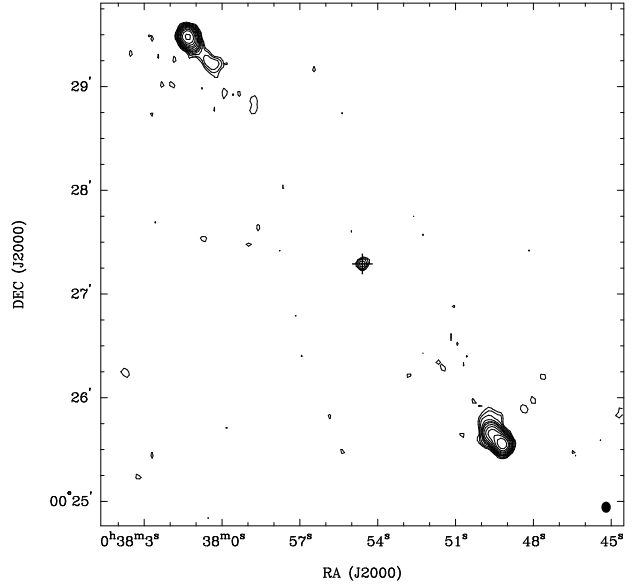}
\caption{The 2SLAQ radio galaxy J003754.59$+$002717.4 at $z$=0.59, which has a projected 
linear size of 2.0\,Mpc (making it one of the largest--known radio galaxies).  A cross 
marks the optical position of the 2SLAQ galaxy, which is coincident with the radio core 
near the centre of the image, and the filled ellipse in the lower right--hand corner shows 
the size of the FIRST beam.   }
\label{J0037.first}
\end{figure} 

We checked all the candidate NVSS identifications by visually inspecting overlays of 
FIRST and NVSS radio contours onto optical images of each galaxy (since the FIRST contours 
usually allow us to unambiguously identify the host galaxy of a candidate NVSS radio source). 
This visual inspection yielded a list of 322 2SLAQ LRGs which appeared to be genuinely 
associated with one or more NVSS radio sources.  271 of these galaxies were already 
identified with FIRST radio sources; but the remaining 51 galaxies had no corresponding 
source in the FIRST catalogue, and are discussed in \S3.3.2 below.  

The NVSS matching process allowed us to identify additional radio components, 
offset more than 30\,arcsec from the 2SLAQ galaxy position, for several 2SLAQ 
LRGs which had already been matched with FIRST sources.  One of these, 
J003754.59$+$002717.4, is shown in Figure~\ref{J0037.first} and is a newly-discovered 
giant radio galaxy with a linear size of 2.0\,Mpc.  Only the relatively weak central 
component of this source  was identified in our original FIRST matching, but the two 
radio hotspots, each more than 2\,arcmin from the 2SLAQ position, were easily found 
by inspection of the NVSS overlay plot. 

\begin{figure}
\vspace*{8cm}
\includegraphics{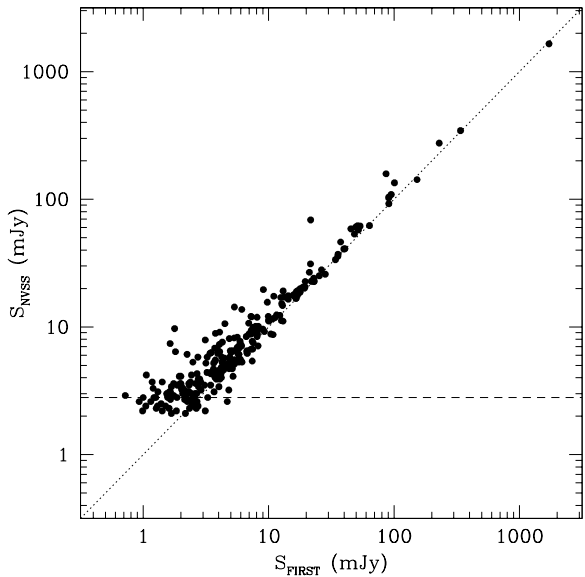}
\caption{Comparison of NVSS and FIRST flux densities for 2SLAQ LRGs detected 
at radio sources. Filled circles show the total flux density of galaxies 
detected in both the FIRST and NVSS catalogues, and the horizontal dashed line 
shows the flux limit of 2.8\,mJy used for calculating the radio luminosity 
function. }
\label{fig.comp}
\end{figure} 

\subsubsection{Comparison of 1.4\,GHz flux densities measured by NVSS and FIRST } 
Figure \ref{fig.comp} compares the total NVSS and FIRST flux densities for 2SLAQ LRGs which 
were detected in both radio surveys.  Note that for sources weaker than about 10\,mJy, the 
catalogued FIRST flux densities tend to be systematically lower than those measured by NVSS, 
mostly due to the radio emission of extended sources being resolved 
out by the smaller FIRST beam (Becker et al.\ 1995; Ivezic et al.\ 2002). 

As noted above, our NVSS cross-matching identified a further 51 candidate NVSS identifications 
of 2SLAQ LRGs which were not matched with radio sources in the FIRST catalogue.  While these 
could be variable radio sources, it is more likely that at the higher resolution of the FIRST 
survey their peak surface brightnesses fall below the $\sim$1\,mJy/beam limit of the FIRST 
catalogue. Three of these objects are associated with wide double sources which are listed 
in the FIRST catalogue but have no catalogued FIRST components within 30\,arcsec of the 
2SLAQ galaxy. 
Many of the remaining objects have a weak radio source at the 2SLAQ 
position which is visible in the FIRST survey image but too faint to be listed in the FIRST 
catalogue.  As noted in \S3.1, the identification of distant LRGs with faint radio sources 
cannot be done reliably using NVSS data alone.  Of the 51 NVSS radio identifications without 
a FIRST catalogue match, we therefore chose to accept as genuine IDs the 22 galaxies which 
had a weak source visible in FIRST images at $>3\sigma$ (where $\sigma$ is the local rms 
noise level in the FIRST image), and within 3\,arcsec of the optical 2SLAQ position.  
This is a conservative approach aimed at maximizing the reliability of our 2SLAQ radio IDs, 
and we recognize that it may have a small effect on the completeness of our radio sample, 
as discussed in \S3.5.2. 

\subsubsection{Statistics of the radio matches}

\begin{table}
\begin{tabular}{|lrrr|}
\hline
Selection & Sample 8 & Sample 9 & All \\
\hline
All objects     & 10072 & 3977 & 14978 \\
Reliable redshift (Q$\geq$3)     &  9307 & 3607 & 13784 \\
Median $z$      &  0.55 &  0.47 & 0.52 \\
Median M$_{0.2,r}$ (mag.) & $-22.54$ & $-22.22$ & $-22.43$ \\
Effective area (deg$^2$) & 141.7 & 93.5 &  \\
FIRST coverage & 98.3\% & 95.5\% & 96.5\% \\
               &       &      &       \\
Radio detections: all&   303 &   71 &  391  \\
Reliable redshift (Q$\geq3$)&   292 &   70 &  378  \\
Radio detection rate &  3.1\%& 1.9\%& 2.7\% \\
\hline
\end{tabular}
\caption{Properties of the 2SLAQ spectroscopic sample used in our analysis and statistics 
of the radio detections.  The calculation of the effective area covered by the survey 
is discussed in \S5.1. 
 }
\label{tab.stats}
\end{table}

Tables \ref{tab.stats} and \ref{tab.comp} summarise the results of matching the 
LRGs in the 2SLAQ  spectrocopic sample with the NVSS and FIRST radio catalogues.  
The overall radio detection rate for LRGs in the spectroscopic sample is close 
to 3\%; but is significantly higher for the galaxies in Sample 8 
($3.1\pm0.2$\%) than for those in Sample 9 ($1.9\pm0.2$\%).  

The difference in radio detection rates is related to the higher median redshift 
of the Sample 8 LRGs compared to those in Sample 9, and a plot of radio detection 
rate versus redshift 
shows a smooth rise in detection rate from about 1.4\% at $z=0.4$ to 3.6\% at 
$z=0.6$ and above.  The detection rate increases with redshift because 
radio galaxies are preferentially found in the most massive (and luminous) 
LRGs (see \S4.2 of this paper). 
Because of the $i<19.8$ apparent magnitude limit of the 2SLAQ sample, these 
very luminous LRGs make up an higher fraction of the 2SLAQ galaxies as the 
redshift increases.  

%



\begin{table}
\begin{tabular}{|lrr|}
\hline
No. of FIRST & No. of & \% of total \\
components  & galaxies & \\
\hline
Uncatalogued    &  22 &  6\% \\
1      & 317 & 81\% \\
2      &  36 &  9\% \\
3      &   9 &  2\% \\
4      &   3 &  1\%\\
5+     &   4 &  1\% \\
Total  & 391 &  \\
\hline
\end{tabular}
\caption{The fraction of radio detections from the 2SLAQ spectroscopic catalogue which 
have more than one FIRST component. Galaxies listed as `uncatalogued' have a radio source 
catalogued in NVSS and a weak source which is visible on the FIRST image but below the 
flux--density limit of the FIRST catalogue. }
\label{tab.comp}
\end{table}

\subsection{The final radio data table}

Table 3 lists the 391 2SLAQ galaxies identified with catalogued FIRST and/or NVSS 
radio sources. 
The columns of Table 3 are as follows: 
\begin{itemize}
\item[(1)]
The 2SLAQ name, set by the J2000 co-ordinates of each galaxy.  
\item[(2)]
The 2SLAQ sample to which the galaxy belongs (see Figure \ref{fig.phot} and Cannon et al.\ 2006). 
\item[(3)]
The dereddened $i$-band ($i_{deV}$) apparent magnitude from SDSS photometry.  
The 2SLAQ spectroscopic LRG catalogue has a magnitude cutoff of $i\leq19.8$\,mag. 
\item[(4)]
The optical position (J2000.0) at which the 2dF fibre was placed. 
\item[(5)]
The radio position, taken from the FIRST catalogue for single sources. For 
radio sources with more than one FIRST component, the position given is the 
radio core (if visible) or the flux--weighted radio centroid. 
\item[(6)]
The peak radio flux density at 1.4\,GHz (in mJy), from the FIRST 
catalogue (Becker et al.\ 1995). 
\item[(7)]
The total 1.4\,GHz flux density from the FIRST catalogue.  
For extended and multiple radio sources with more than one 
entry in the FIRST catalogue, the flux density quoted here 
is the sum of all the components.
\item[(8)]
The number of FIRST radio components associated with each 
2SLAQ galaxy.
\item[(9)]
The total 1.4\,GHz flux density from the NVSS catalogue.  
For radio sources with more than one NVSS component, the flux density quoted here 
is the sum of all the components.
A handful of sources had a single NVSS component and two FIRST components, only one 
of which is genuinely associated with the 2SLAQ galaxy. In this case, the listed NVSS 
flux density of these sources was reduced in proportion to the flux--density ratio 
of the associated and unassociated FIRST components.  
\item[(10)]
The error on the total NVSS flux density. 
\item[(11)]
The number of NVSS radio components associated with each 
2SLAQ galaxy.
\item[(12)]
The heliocentric redshift $z$ measured from 2SLAQ optical spectra. 
\item[(13)]
The 2dF redshift quality code Q, where Q=4 or 5 are excellent--quality 
redshifts ($>$99\% reliable), Q=3 is acceptable ($>$95\% reliable) and Q=0, 1 
or 2 indicates a poor--quality or highly-unreliable redshift.  Galaxies with 
Q$<$3 were excluded from further analysis because their redshifts are 
highly uncertain. 
\item[(14)] The extinction--corrected $r$--band absolute magnitude M$_{0.2,r}$ 
for each galaxy, calculated using the method described by Wake et al.\ (2006).  
This incorporates both a $k$-correction and a further correction for the passive 
evolution of the stellar population, and represents the absolute magnitude 
which each LRG would have if observed through a standard SDSS $r$-band filter 
redshifted to $z=0.2$. 
\item[(15)] The total radio luminosity at 1.4\,GHz, calculated by assuming a mean 
spectral index $\alpha=-0.7$ (where flux density S$_\nu\propto\nu^\alpha$) and the 
usual $k$-correction of the form $k(z)=(1+z)^{-(1+\alpha)}$. 
\item[(16)] Notes on individual galaxies.  
\end{itemize}

\begin{figure}
\vspace*{8.5cm}
%
%
\includegraphics{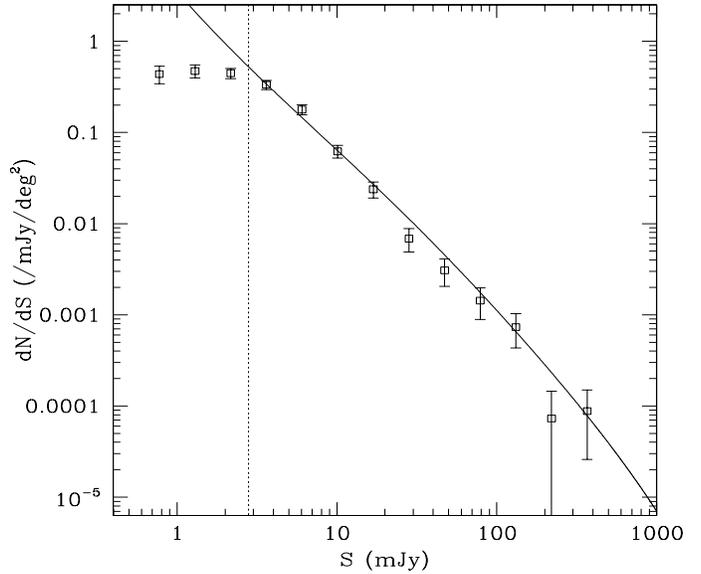}
\caption{Log N -- log S plot for the final 2SLAQ radio sample at 1.4 GHz.  
Open squares show the observed flux--density distribution, and the vertical dashed 
line indicates the flux density limit of 2.8\,mJy used in calculating the radio 
luminosity function (see \S5).  The solid line shows the overall 1.4\,GHz source 
counts from Hopkins et al.\ (2003), scaled by an arbitrary factor of 0.05 in dN/dS 
for ease of comparison.  Note that the values of dN/dS in this plot are differential 
source counts, and we chose not to follow the radio convention of dividing by a 
Euclidean slope of S$^{-2.5}$. }
\label{logn.logs} 
\end{figure} 

\subsection{Reliability and completeness of the final radio sample} 
\subsubsection{Reliability} 
The great majority of the galaxies in Table 3 (81\%) are associated with single--component 
FIRST sources, and Monte Carlo tests show that these can be identified with a high 
degree of reliability (fewer than 1\% will be chance associations of unrelated objects). 
For galaxies with multiple FIRST components, which are accepted on the basis on visual 
inspection, the reliability should also be high (Best et al.\ (2005a) assumed their 
visual analysis to be 100\% reliable).  We therefore estimate that the reliability of 
our final 2SLAQ radio sample is at least 98\%, similar to the overall reliability 
of 98.9\% quoted by Best et al.\ (2005a) for their lower--redshift SDSS sample. 

\subsubsection{Completeness} 
In assessing the overall completeness of our final radio sample, we need to examine  
the completeness of both the optical and the radio samples. 
Only a subset of the 2SLAQ LRG photometric catalogue was observed spectroscopically, 
but we can account for this by calculating the effective area of sky covered by the 
2dF spectroscopic sample as described in \S5.1. For the luminosity function 
calculations presented in \S5, the only remaining incompleteness on the optical side comes 
from the small fraction of galaxies for which the measured redshift is unreliable 
(i.e. objects with Q$\leq$2 in Table 3). Of the radio--detected LRGs in Table 1, 
378/391 have a reliable 2dF redshift so the spectroscopic incompleteness of our sample 
is about 3\%. 

On the radio side, the NVSS catalogue is essentially complete for radio sources with 
flux densities above 2.8\,mJy, as can be seen from Figure 32 of Condon et al.\ (1998).  
This is supported by the number counts of 2SLAQ radio sources shown in Figure \ref{logn.logs}, 
which account for roughly 5\% of all 1.4\,GHz radio sources in the flux--density range 
2.8--500\,mJy based on the source counts presented by Hopkins et al.\ (2003).  
Below the NVSS completeness limit of 2.8\,mJy the number counts in Figure \ref{logn.logs} 
start to fall off and our radio sample becomes seriously incomplete, reflecting the 
well-known incompleteness of the FIRST catalogue for faint, extended radio sources.  


Our decision to reject candidate NVSS IDs for which we could find no corresponding source at 
the 3$\sigma$ level in the FIRST images (see \S 3.3.2) may also affect the completeness level 
of our final sample.  If all 15 of the rejected NVSS objects with flux density above 2.8\,mJy 
are actually genuine IDs and should have been included in our sample, this would represent 
a radio incompleteness of 2\%.   

As discussed in \S5, we set a radio flux--density limit of 2.8\,mJy in our 
calculation of the radio luminosity function.  If the NVSS catalogue is 
essentially complete above this limit, 
then the overall incompleteness of our 2SLAQ radio sample is roughly 5\% (i.e. 3\% spectroscopic 
incompleteness, plus up to 2\% incompleteness in the radio catalogue) for sources with 
radio flux densities above 2.8\,mJy. 
We therefore estimate that our sample of 2SLAQ radio galaxies is at least 95\% complete and 
98\% reliable down to the magnitude and flux--density limits used for the luminosity function 
calculations in \S5. 

\section{Characteristics of the LRG radio--galaxy population } 

\subsection{Physical mechanisms for the radio emission} 
In the 2SLAQ redshift range ($0.4<z<0.8$), even the weakest radio sources 
in the FIRST and NVSS catalogues will have radio luminosities above 10$^{24}$\,W\,Hz$^{-1}$. 
In the local universe, radio sources above this level are almost invariably associated 
with active galactic nuclei (AGN) rather than star--forming galaxies (Condon 1992; 
Sadler et al.\ 2002; Best et al.\ 2005a).  Since the photometric selection of the 2SLAQ 
LRG sample also excludes galaxies with ongoing star formation, we expect all the radio 
galaxies identified in Table 3 to be powered by an AGN. 

\setcounter{table}{3}
\begin{table}
\begin{tabular}{|lrrr|}
\hline
\multicolumn{1}{c}{Absolute mag.} & \multicolumn{1}{c}{All 2SLAQ}  & \multicolumn{1}{c}{Radio-det.}  & \multicolumn{1}{c}{Radio det.} \\
\multicolumn{1}{c}{M$_{0.2,r}$} & \multicolumn{1}{c}{LRGs}  & \multicolumn{1}{c}{LRGs}  & \multicolumn{1}{c}{rate (\%)} \\
\hline
   $-20.25$ to $-20.75$ & 8      &    0    &   0.0  \\
   $-20.75$ to $-21.25$ & 71     &    0    &   0.0  \\
   $-21.25$ to $-21.75$ & 514    &    1    &   0.2  \\
   $-21.75$ to $-22.25$ & 2870   &   26    &   0.9  \\
   $-22.25$ to $-22.75$ & 4271   &  143    &   3.3  \\
   $-22.75$ to $-23.25$ & 2063   &  140    &   6.8  \\
   $-23.25$ to $-23.75$ & 429    &   54    &  12.6  \\
   $-23.75$ to $-24.25$ & 35     &    9    &  25.7  \\
   $-24.25$ to $-24.75$ & 4      &    2    &  50.0  \\
      Total   & 10265  &  375    &         \\
        & & & \\
Median M$_{0.2,r}$ & \multicolumn{1}{r}{$-22.43$} & \multicolumn{1}{r}{$-22.81$} & \\
                   & \multicolumn{1}{r}{(3.7\,L$^*$)} & \multicolumn{1}{r}{(5.3\,L$^*$)} & \\
\hline
\end{tabular}
\caption{Distribution of the full 2SLAQ LRG sample, and the radio--detected subsample, 
in 0.5\,mag.\ bins in $r$-band absolute magnitude M$_{0.2,r}$. Only galaxies with reliable 
(Q$\geq$3) redshifts in the range $0.3<z<0.8$ are included in this table.  }
\label{tab.magstats}
\end{table}

\subsection{Redshift and luminosity distribution} 
The radio galaxies in the 2SLAQ LRG sample are significantly more luminous than the  
2SLAQ LRG sample as a whole, as can be seen from Table \ref{tab.magstats}.  This is not 
unexpected, since the probability that an early--type galaxy will be a radio galaxy 
is known to rise sharply with optical luminosity (Auriemma et al.\ 1977; 
Sadler, Jenkins \& Kotanyi 1989, Best et al.\ 2005b).  

\begin{figure}
\vspace*{8.5cm}
\includegraphics{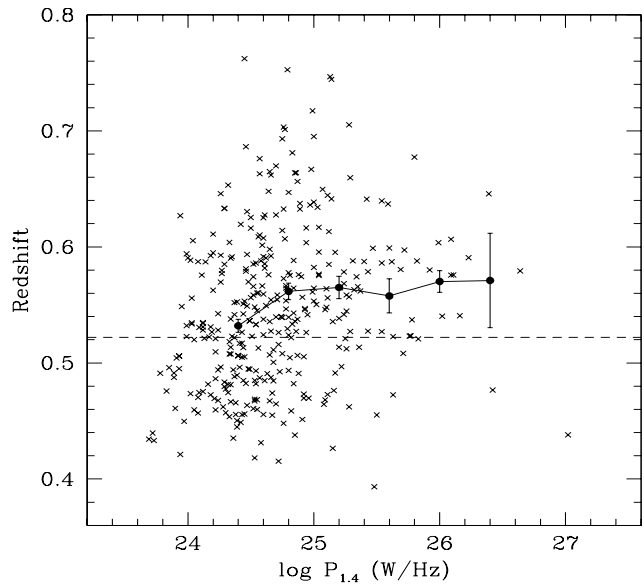}
\caption{The mean redshift of 2SLAQ radio galaxies as a function of radio power.
Crosses show individual radio sources, while the 
filled circles and solid line show the mean redshift for each of the 
0.4 dex bins in log P used to calculate the radio luminosity function. 
The error bars show the standard error on the mean for each bin, and the 
horizontal dashed line is the mean redshift for the full 2SLAQ LRG 
spectroscopic sample.  }
\label{fig.zlum}
\end{figure}

\begin{figure}
\vspace*{8.5cm}
%
%
\includegraphics{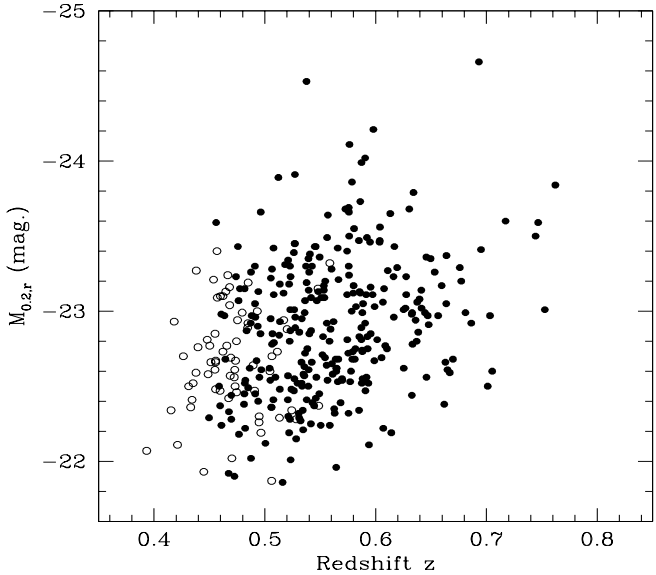}
\caption{Distribution of $r$-band absolute magnitudes for 2SLAQ radio galaxies over 
the redshift range of the sample.  The absolute magnitudes M$_{0.2,r}$ are those 
defined by Wake et al.\ (2006), and represent the absolute magnitude of 
an LRG as it would appear if redshifted to $z=0.2$ and observed through 
an SDSS $r$-band filter. 
Filled circles show galaxies from Sample 8 (see Figure 1) and open circles 
galaxies from Sample 9. 
}
\label{fig.absmag} 
\end{figure}

Figures \ref{fig.zlum} and \ref{fig.absmag} show that the 2SLAQ radio sample is  
close to volume--limited, with no strong correlation between either absolute 
magnitude or radio luminosity and redshift.  In particular, Figure \ref{fig.zlum} 
shows that the mean redshift of the 2SLAQ radio galaxies is almost independent 
of their radio luminosity, making this an excellent sample for measuring the radio 
luminosity function.  The smaller number of low--power ($<10^{24.8}$\,W\,Hz$^{-1}$) 
radio galaxies at redshifts above $z\sim0.6$ is due to the flux--density limit of the 
NVSS and FIRST catalogues, and causes the mean redshift to drop slightly at the 
lowest radio luminosities.  This flux limit is taken into account when calculating 
the radio luminosity function, and so does not affect the sample completeness. 

Figure \ref{fig.absmag} shows the relatively sharp cutoff in absolute magnitude M$_{0.2,r}$ 
caused by the $i<19.8$\,mag.\ cutoff of the 2SLAQ LRG sample.  Although the 2SLAQ LRGs 
span a wide range (almost three orders of magnitude) in radio luminosity, their range 
in optical luminosity is much narrower, corresponding to only one order of magnitude 
(roughly 2--20\,L$^*$). 

Since Wake et al.\ (2006) find no evolution in 
the number density of these bright LRGs over the redshift range $0<z<0.6$ beyond 
that expected from the passive evolution of their stellar population, it is 
straightforward to match the 2SLAQ sample with local early--type galaxies of 
similar luminosity in order to study the evolution of their radio properties. 

\subsection{Optical emission lines} 

\subsubsection{The fraction of galaxies with optical emission lines}
We visually inspected all the spectra of radio--detected 2SLAQ galaxies 
to check for the presence of optical emission lines.  Where seen, 
this is noted in Table 3.  In most cases the only detected emission 
line is [O\,II] at 3727\AA, which is seen in 27\% of 2SLAQ radio galaxies. 
Thus, like most low--power radio 
galaxies in the local universe (Owen, Ledlow \& Keel 1995), the majority 
of 2SLAQ radio galaxies do not show strong optical emission lines and would 
not have been recognized as AGN on the basis of their optical spectra alone. 

\subsubsection{[O\,II] emission} 
The 3727\AA\ [O\,II] emission line can be excited both by star--forming regions 
and by an AGN; 
but it can also occur when neither of these is present, through photoionization 
by the old, hot post--AGB stars which should exist in all luminous early--type 
galaxies (Bressan et al.\ 1994).  As a result, detection of the [O\,II] line 
alone gives little or no useful information about either the ionization 
state of the gas or the dominant physical process responsible for ionizing it.  

Measurement of the observed ratios of other common emission lines such as H$\alpha$, 
H$\beta$, [O\,III] and [N\,II] (e.g. Veillueux \& Osterbrock 1987) can distinguish 
star--forming galaxies from those in which the gas is ionized by an AGN, but such 
measurements are not possible for most of the LRGs in our 2SLAQ sample because the 
relevant emission lines are redshifted out of the 2dF spectral range. 

\subsubsection{[Ne\,III] and [Ne\,V] emission lines}
Thirteen of the galaxies in Table 3 (3.4\% of the 
radio sample) show both strong [O\,II] emission lines and weaker emission lines 
of [Ne\,III] 3869\,\AA\ and/or [Ne\,V] 3426\,\AA, as seen in Figure 2(c). 
The presence of these additional high--excitation emission lines strongly suggests 
that the dominant ionizing source in these particular galaxies is an active nucleus. 

The fraction of radio--detected LRGs with detected [Ne\,III] and [Ne\,V] emission 
lines, though low, is significantly higher than in the 2SLAQ LRG sample as a whole. 
We estimate that only $0.4\pm0.2\%$ of a sample of 
744 good--quality 2SLAQ LRG spectra show visible [Ne\,III] and [Ne\,V] emission lines, 
so the line is roughly eight times more common in the radio--detected subsample 
than in the LRG population as a whole. We stress once again, however, that the 
majority of the low--power radio galaxies in the 2SLAQ sample are optically 
`normal' and do not show strong optical emission lines in their spectra. 

\subsubsection{Comparison with the full 2SLAQ LRG sample}
As noted by Roseboom et al.\ (2006), the relatively low S/N of the 2SLAQ spectra  
makes it difficult to measure the equivalent widths of weak emission lines accurately. 
They therefore used a cutoff in [O\,II] emission--line equivalent width of 
$\geq$8\,\AA\ to define an emission--line LRG, and found that 15\% of the 2SLAQ 
LRGs lay above this cutoff. 

Our visual classification of emission--line 
galaxies in Table 3 selects objects with almost exactly the same range in [O\,II] 
equivalent width as the `em'  and `em+a' galaxies discussed by Roseboom et al.\ (2006). 
For a sample of 162 galaxies in common, 24/26 (92\%) of the 
Roseboom et al.\ (2006) objects with [O\,II] EW$\geq$8.0\AA\ are listed by us 
as having [O\,II] emission, compared to only 11/136 (8\%) of the EW $<$8.0\AA\ 
objects.  If we drop the equivalent width cutoff to 7.0\,\AA, the agreement is 
even better, with 93\% of galaxies above the cutoff and only 5\% of those below 
the cutoff classifed in Table 3 as showing [O\,II] emission.  

The fraction of galaxies with [O\,II] emission in the full sample of 5,697 2SLAQ 
LRGs measured by Roseboom et al.\ (2006) is 15.1\% if we adopt an EW cutoff of 
8\,\AA, and 17.7\% for an EW cutoff of 7\,\AA.  This is lower than the 
27\% [O\,II] detection rate for the 2SLAQ radio galaxies in Table 3, but the 
difference needs to be assessed carefully.  The 
emission--line luminosity of elliptical and S0 galaxies correlates with absolute magnitude, 
in the sense that luminous galaxies are more likely than smaller galaxies to 
show emission lines in their spectra (Phillips et al.\ 1986). 
To compare the emission--line properties of 2SLAQ radio galaxies with those of the 
full LRG sample therefore, we need to compare sets of objects which are carefully 
matched in optical luminosity. Such a comparison is outside the scope of the present 
paper, but will be included in a later paper in this series. 



\section{The radio luminosity function at of Luminous Red Galaxies at z=0.4 to 0.8}

\subsection{Area covered by the 2SLAQ spectroscopic survey} 
As discussed by Cannon et al.\ (2006), the spectroscopic coverage 
of the 2SLAQ LRG sample is not uniform and has both overlapping fields 
and gaps on the sky. Determining the total area of sky covered by the 2SLAQ 
spectroscopic sample is therefore a necessary first step in measuring the 
radio luminosity function. 

Cannon et al.\ (2006) estimated the effective area of the 2SLAQ LRG 
spectroscopic survey as $\sim$135\,deg$^2$ for Sample 8 and 
$\sim$90\,deg$^2$ for Sample 9, 
based on the number of galaxies observed and their surface density 
on the sky.  
Since an accurate measurement of effective area is crucial to the 
normalization of the measured radio luminosity function, we have 
recalculated the 2SLAQ LRG survey area using the same methodology 
applied by Wake et al.\ (2006) to calculate the optical luminosity 
function for the 2SLAQ LRGs. 

Both the completeness and the area were calculated using a mask which 
is constructed by repeatedly running the 2dF `configure' program on a random 
distribution of points in one field. This creates a random catalogue of 
roughly 5 million points, with a distribution covering all the possible 
positionings of a 2dF fibre. A random catalogue for the entire 
survey was then constructed by
placing this single random 2df field distribution at every field centre 
which was observed in the survey.  
Regions which are not in the input catalogue were removed, since the 
edges of some of our fields lay outside the region covered by the input
catalogue.  This final random catalogue contained about 400 million data 
points. 

The final mask was based on a grid with 30$\times$30\,arcsec pixels 
covering the whole survey area. Those pixels which contained at least one 
random point were flagged, and the mask was used to calculate both the 
area and the completeness. To calculate the area, we simply summed the pixels 
which were flagged as containing a random point; this yielded a total area of 
186.2\,deg$^2$.  
To determine the completeness, we divided the number of LRGs with a redshift in
the masked region by the number of LRGs from the input catalogue in the
masked region. 
This allows calculation of the effective area for each spectroscopic 
sample, which is simply the total area multiplied by the completeness.  
For galaxies with spectra whose quality is good enough to measure a reliable 
redshift (Q$\geq$3), we calculate an effective area of 141.7\,deg$^2$ for 
Sample 8 and 93.5\,deg$^2$ for Sample 9.  These are roughly 4--5\% higher than 
the values estimated by Cannon et al.\ (2006).  

In this study, we must also account for the small fraction of the 2SLAQ survey area 
(mostly at low Galactic latitude) for which radio data from the FIRST survey are not 
available. 
For the 2SLAQ spectroscopic sample, 1.7\% of galaxies in Sample 8 and 4.5\% of galaxies 
in Sample 9 lie outside the region of sky covered by the FIRST source catalogue. 
The smaller overlap between FIRST and Sample 9 is because 
more of the Sample 9 fields are at lower galactic latitude than the region covered by FIRST. 
After accounting for the area not covered by FIRST, we have a final effective area 
of 139.3\,deg$^2$ and 89.3\,deg$^2$ for Samples 8 and 9 respectively. 

\subsection{Calculation of the radio luminosity function} 
We derived the radio luminosity function for 2SLAQ LRGs by applying the $1/V_{\rm max}$ 
method (Schmidt 1968) to the 251 2SLAQ LRGs which belong to Sample 8 or 9 and 
have a 1.4\,GHz flux density of 2.8\,mJy or more.  We calculated $V_{\rm max}$ for 
each object using the maximum and minimum redshift limits at which each object 
could be placed and still satisfy both the optical and radio selection criteria of our 
sample.  The upper and lower redshift limits set by the optical colour and magnitude 
cutoffs for the 2SLAQ sample 
were calculated using the same method as Wake et al.\ (2006) for their calculation of 
the optical luminosity function, while the radio upper limits were set by the redshift 
at which each source fell below the 2.8\,mJy flux limit (there is no lower redshift limit 
for our radio selection criteria). The value of $V_{{\rm max},i}$ adopted for the $i$th 
sample galaxy is: 

\begin{equation}
V_{{\rm max},i}=\frac{\Omega}{4\pi}\left(\min\{V_{{\rm max},i}^{\rm optical},V_{{\rm max},i}^{\rm radio}\} - V_{{\rm min},i}^{\rm optical}\right),
\end{equation}

\noindent
where $\Omega$ is the effective area of samples 8 and 9 in steradians as listed in 
Table \ref{tab.stats}, and $V_{{\rm max},i}^{\rm optical},V_{{\rm max},i}^{\rm radio}$ 
and $V_{{\rm min},i}^{\rm optical}$ 
are the (all-sky) volumes enclosed by upper and lower redshift limits 
$z_{\rm max}$ and $z_{\rm min}$.  Table 5 lists the radio luminosity function which we 
measured for 2SLAQ LRGs in the redshift range $0.4<z<0.8$. 

\begin{table*}
\centering
\begin{minipage}{140mm} 
\caption{The radio luminosity function at 1.4\,GHz, for local LRGs from the 
6dF Galaxy Survey (Mauch \& Sadler 2006), and for 2SLAQ LRGs at $z\sim0.55$ from this paper. 
In both cases the mean value of $\langle$V/V$_{\rm max}\rangle$ is close to 0.5, 
consistent with the two samples being essentially complete and volume--limited.   }
\begin{tabular}{@{}crcrc@{}}
\hline
   & \multicolumn{2}{c}{Local AGN: M$_K<-24.8$} & \multicolumn{2}{c}{2SLAQ LRG } \\ 
$\log_{10}$L$_{1.4}$ & N & \multicolumn{1}{c}{$\log_{10}\Phi$} & N & 
\multicolumn{1}{c}{$\log_{10}\Phi$} \\ 
 (W Hz$^{-1}$) &   & (mag$^{-1}$ Mpc$^{-3}$)& & (mag$^{-1}$ Mpc$^{-3}$) \\
\hline
   23.6    & 328 & $-5.23^{+0.03}_{-0.03}$ &     & \\
   24.0    & 255 & $-5.39^{+0.03}_{-0.03}$ &     & \\
   24.4    & 178 & $-5.62^{+0.04}_{-0.04}$ &  61 &  $-5.45^{+0.08}_{-0.10}$  \\
   24.8    &  80 & $-6.05^{+0.06}_{-0.07}$ & 102 &  $-5.73^{+0.05}_{-0.06}$  \\
   25.2    &  49 & $-6.33^{+0.08}_{-0.09}$ &  52 &  $-6.17^{+0.06}_{-0.07}$  \\
   25.6    &  15 & $-6.97^{+0.12}_{-0.16}$ &  22 &  $-6.43^{+0.10}_{-0.13}$  \\
   26.0    &   3 & $-7.29^{+0.25}_{-0.63}$ &   9 &  $-6.93^{+0.13}_{-0.19}$  \\
   26.4    &   2 & $-8.64^{+0.25}_{-0.67}$ &   3 &  $-6.84^{+0.26}_{-0.74}$  \\
   26.8    &     &                         &   1 &  $-7.73^{+0.30}_{-1.00}$  \\
   27.2    &     &                         &   1 &  $-8.09^{+0.30}_{-1.00}$  \\
&&&& \\
& $\langle$V/V$_{\rm max}\rangle$ &  
\multicolumn{1}{c}{0.519$\pm$0.006} & 
 & \multicolumn{1}{c}{0.455$\pm$0.018 } 
 \\
\hline
&&&& \\
\end{tabular}
\end{minipage}
\end{table*}

\subsection{Evolution of the radio--galaxy population to $z\sim$0.7}
\subsubsection{Comparison with the local radio--galaxy population}  
We can now determine the evolution of the radio--galaxy population out to $z\sim0.7$ by 
comparing the 2SLAQ radio luminosity function with the RLF of similar galaxies 
in the local universe.  For our local benchmark, we need a large, uniform survey which 
covers a large volume at $z\sim0$ (so that evolutionary effects within the sample volume 
can be neglected). 

We chose to use as our comparison sample the large dataset of nearby radio galaxies 
identified by Mauch \& Sadler (2006) from the second data release of the 6dF Galaxy Survey 
(6dFGS; Jones et al.\ 2004).  The 6dFGS DR2 contains spectra of about 45,000 galaxies brighter 
than K=12.75\,mag. in the near--infrared K--band, and the 6dFGS radio galaxies identified by 
Mauch \& Sadler span the redshift range $0<z<0.15$ with a median redshift of 0.073.  
The large sample volume ($\sim3.9\times10^8$\,Mpc$^3$) and shallow depth of the 6dFGS survey 
allow us to derive an accurate $z\sim0$ radio luminosity function with which to compare our 
2SLAQ results. 
 
We note that passive evolution of the optical luminosity function of 
luminous red galaxies over the redshift range $0<z<0.7$ has little effect  
on the results of this study, since radio galaxies are known to be almost 
exclusively associated with the most massive and luminous galaxies 
at all redshifts out to $z\sim5$ (Rocca--Volmerange et al.\ 2004).  
However, we have corrected the optical luminosities of the 2SLAQ galaxies 
to their equivalent value at $z=0.2$ 
using the same correction factors for passive evolution as Wake et 
al.\ (2006)  The values of M$_{0.2,r}$ for 2SLAQ LRGs in this paper should differ by 
less than 0.1\,mag from the M(r) values measured for similar galaxies at $z=0$.

Since the 2SLAQ LRG sample effectively excludes galaxies with absolute magnitudes 
fainter than M$_{0.2,r}\sim-22.0$, we also chose to compare the 2SLAQ RLF with a 
local RLF calculated for a sample with a similar cutoff in optical luminosity. 
The optical luminosity function for 6dFGS galaxies (Jones et al.\ 2006) has 
M$_r^*$=$-$21.0\,mag.\ and M$_{\rm K}^*$=$-$23.8\,mag., so we adopted a cutoff in 
absolute magnitude of M$_{\rm K}<-24.8$\,mag.\ 
(i.e. one magnitude brighter than M$_{\rm K}^*$) for our local sample in order to 
match the luminosity cutoff of the 2SLAQ sample as closely as possible.  
In practice, since the probability that an early--type galaxy will be a radio galaxy 
rises very sharply with optical luminosity, 
this luminosity cut has almost no effect on the measured RLF for local galaxies --- 
only six of the 337 6dFGS radio galaxies with radio luminosities P$_{1.4}>10^{24.2}$\,W\,Hz$^{-1}$ 
are excluded by our cutoff in M$_{\rm K}$. 

\begin{figure}
\vspace*{8.5cm}
\includegraphics{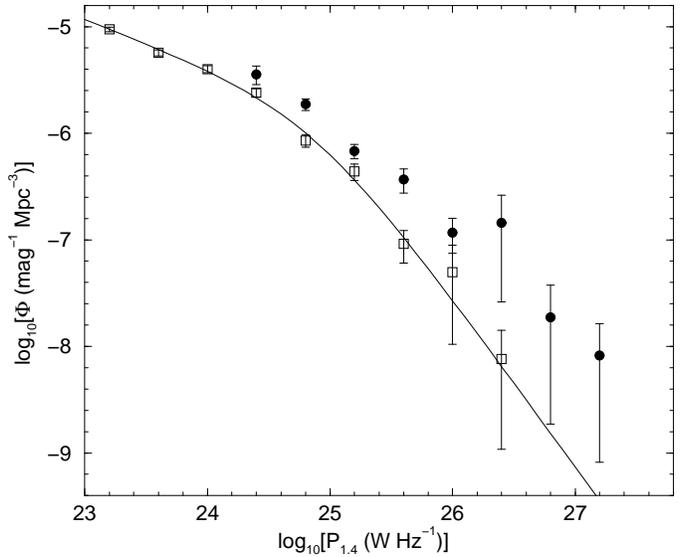}
\caption{The local radio luminosity function (RLF) for 2SLAQ LRGs (filled circles), 
compared with the observed local RLF measured by Mauch \& Sadler (2006) for luminous 
radio galaxies (M$_{\rm K}<-24.8$\, mag.) in the 6dF Galaxy Survey (open squares). 
The dashed line shows a parametric fit to the 6dFGS RLF over the luminosity range 
10$^{22}$ to 10$^{26}$\,W\,Hz$^{-1}$.  }
\label{fig.rlf} 
\end{figure} 

\begin{figure}
\vspace*{8.5cm}
\includegraphics{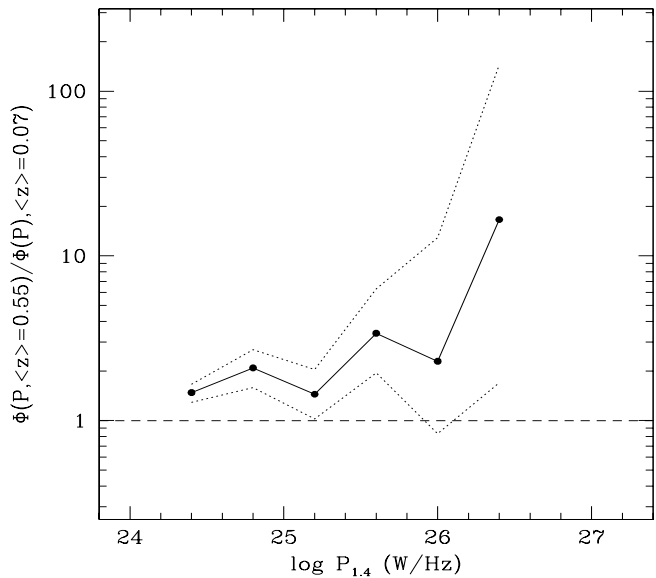}
\caption{Ratio of the radio luminosity functions measured from the 2SLAQ 
(median redshift $z$ = 0.55) and 6dFGS (median redshift $z$ = 0.07) samples. 
Dotted lines show the range of possible values set by the error bars on the two 
RLFs.  }
\label{fig.evol1} 
\end{figure}

Figure \ref{fig.rlf} compares the 6dFGS and 2SLAQ RLFs for the range in radio luminosity 
for which they overlap.  The 2SLAQ RLF is higher than the local value at all luminosities, 
but the difference is most striking for the most powerful radio galaxies in our sample.  
This can be seen more clearly in Figure \ref{fig.evol1}, which plots the ratio of the 
space densities of 2SLAQ and 6dFGS radio galaxies.  
Figure \ref{fig.evol1} strongly suggests that  low--power radio galaxies 
undergo cosmic evolution, though their evolution is less dramatic than that seen 
in powerful radio galaxies.

\subsubsection{The redshift evolution of low--power radio galaxies} 

The redshift evolution of a galaxy population can be investigated by comparing the 
measured luminosity function of the same population at two or more epochs.  
Any shift in the luminosity function is usually represented 
by one of two simple extremes: {\sl luminosity evolution}, in which the luminosity 
functions are matched by shifting one of them horizontally; and {\sl density evolution}, 
in which the shift is vertical.  These alternatives are equally plausible for a 
pure power--law luminosity function, but can in principle be distinguished if the 
luminosity function undergoes a change in slope at some luminosity. 

As noted by Peacock (1999) the physical motivation for either of these descriptions 
is weak.  Both luminosity and density evolution assume 
that the overall shape of the luminosity function remains the same at all epochs, 
which may not be the case. Luminosity evolution suggests a population 
whose overall luminosity declines with time, while density evolution implies a set 
of objects which have constant luminosity but a range of lifetimes.  It is likely 
that neither of these simple models completely describes the physical processes which 
actually occur.  Our main goal 
in this paper, however, is to determine whether low--power radio galaxies 
evolve with redshift and even a simple parameterization of the luminosity function is 
an adequate tool for testing this. 

\begin{figure}
\vspace*{8.5cm}
%
\includegraphics{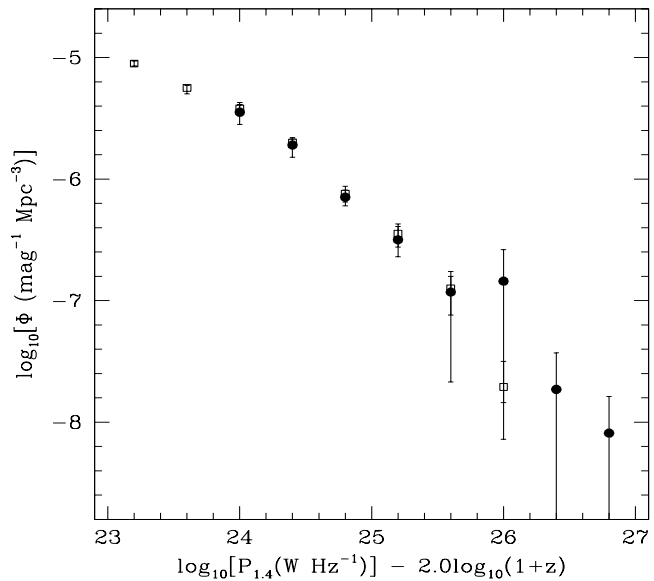}
\caption{The predicted radio luminosity functions of 2SLAQ LRGs (open circles) and 6dFGS galaxies 
(filled circles) at the current epoch, where the equivalent luminosity function at $z$=0 
has been derived by dividing the observed luminosity function of each sample by 
(1+$z$)$^{2.0}$.  }
\label{fig.evol2} 
\end{figure}

To quantify the evolution of low--power radio galaxies, we follow earlier studies 
in assuming that the redshift dependance can be represented by pure 
luminosity evolution of the form 
$$P^*(z)=P^*(0).(1+z)^{K_L}$$
(Boyle et al.\ 1988). 
We then find the best-fitting value for $K_L$ by dividing the radio power
of each galaxy by $(1+z)^{K_L}$ for $0<K_L<10$ and minimising the
$\chi^2$ value from the rebinned local and 2SLAQ RLFs for each $K_L$.  
Error estimates are determined by computing the value of $K_L$ for which
$\chi^2-\chi^2_{\rm min}=1$, which is equivalent to $1\sigma$
(Lampton, Margon, \& Bowyer 1976). The fit of $K_L$ was only made
for $P<10^{25.8}$\,W\,Hz$^{-1}$ to avoid contamination by the more
strongly--evolving high--power radio galaxies in the 2SLAQ sample.
The best fitting value for $K_L$ is $2.0\pm0.3$, which implies that
low--power radio galaxies in the redshift range probed by the 
2SLAQ sample are significantly more luminous than those in the local sample.
The hypothesis that low--power radio galaxies undergo no cosmic evolution can 
be ruled out at the 6.7$\sigma$ level. Figure \ref{fig.evol2} shows the 6dFGS 
and 2SLAQ RLFs shifted to redshift $z$=0 assuming the fitted form for the evolution. 

In summary, we find that low--power radio galaxies with P$_{1.4}<10^{25}$\,W\,Hz$^{-1}$ 
undergo significant cosmic evolution over the redshift range $0<z<0.7$.  Our observations 
are consistent with these low--power radio galaxies undergoing pure luminosity evolution 
of the form (1+$z$)$^k$ where $k=2.0\pm0.3$. 
This is close to the value of 
$k={2.7\pm0.6}$ found by Hopkins (2004) for the luminosity evolution of 
the radio emission from star--forming galaxies over a similar redshift range. 

\subsubsection{The redshift evolution of powerful radio galaxies} 

Although the number of powerful radio galaxies in our 2SLAQ sample is relatively 
small, there is some evidence that they undergo more dramatic evolution than 
their lower--power counterparts.  Figure \ref{fig.evol1} implies that 
the number density of the most powerful (P$_{1.4}>10^{26}$\,W\,Hz$^{-1}$) 
radio galaxies in our sample could be at least ten times higher at $z\sim0.55$ 
than in the local universe, though this is highly uncertain because of the small number 
of objects observed in this luminosity range.  
The 6dFGS radio--galaxy sample covers 
a volume of $3.9\times10^8$\,Mpc$^3$, i.e. almost four times the 2SLAQ volume, but 
has only four galaxies with radio luminosity above 10$^{26}$\,W\,Hz$^{-1}$, whereas 
the smaller 2SLAQ volume contains at least ten radio galaxies more powerful than 
$10^{26}$\,W\,Hz$^{-1}$.  We therefore stress that the small number of powerful 
radio galaxies seen at low redshift is not a volume effect.  

If we assume the same luminosity evolution of the form (1+$z$)$^2$  which fits 
the low--power sample, then the observation of four P$_{1.4}>10^{26.0}$\,W\,Hz$^{-1}$ 
radio galaxies in the 6dFGS volume implies that we would expect to see 1.23 radio galaxies 
with P$_{1.4}>10^{26.32}$\,W\,Hz$^{-1}$ in the 2SLAQ sample volume.  Five are actually 
observed, so Poisson statistics imply a probability of $\sim$0.7\% that the difference 
is due to chance.  If we also take into account the formal uncertainty in the four 
high--power 6dFGS sources, then the significance level drops to 3--4\%. The small number 
of very powerful radio galaxies in the 6dFGS and 2SLAQ samples therefore limits 
what we can say about their cosmic evolution at this stage. 
We note however that the observed number of powerful ($>10^{26}$\,W\,Hz$^{-1}$) radio 
galaxies in the 2SLAQ sample is consistent with a population undergoing the kind 
of rapid density evolution reported by Willott et al. (2001), who found an increase of 
3\,dex in the space density of the most powerful radio galaxies over the redshift 
range $0<z<2$.  This implies a number density increasing as approximately $(1+z)^6$, 
i.e. ten times as many radio galaxies at powers above 10$^{26}$ in the 2SLAQ sample 
at $z\sim0.55$ as in the 6dFGS at $z\sim0.07$, which is close to the value
seen in Figure \ref{fig.evol1}.

\section{Discussion}

\subsection{Environment of the 2SLAQ radio galaxies} 
As discussed briefly in \S3.2.2, radio--source statistics suggest that many 
of the 2SLAQ galaxies lie in highly--clustered environments. 
The 2SLAQ sample also contains at least two  $z>0.5$ ``head--tail'' radio galaxies 
(J135058.43$-$003633.5 and J143432.65$-$010457.3 in Table 3), which are often 
tracers of high--redshift rich clusters (e.g. Blanton et al.\ 2003). 

\begin{figure}
\vspace*{8.5cm}
\includegraphics{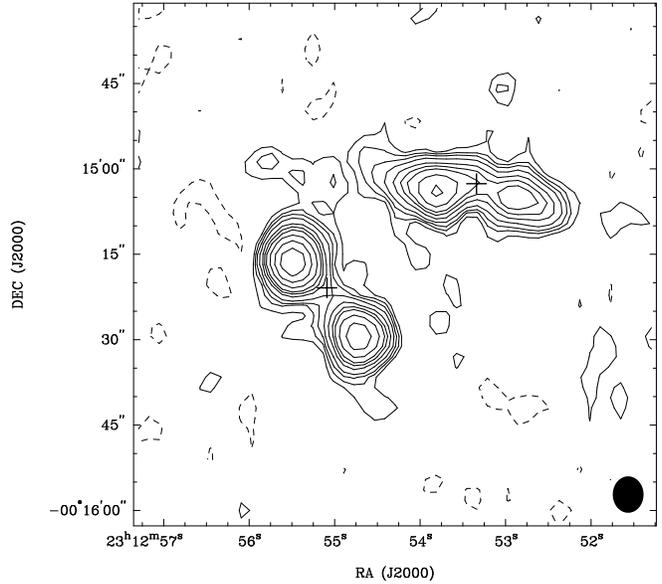}
\caption{FIRST contour plot of the 2SLAQ radio galaxies 
J231253.34$-$001502.5 ($z$=0.5875) and J231255.09$-$001520.8 ($z$=0.5861). 
Crosses mark the positions of the two optical galaxies, and the filled ellipse st lower right 
shows the FIRST beam.  }
\label{fig.cl}
\end{figure}

The clustering properties of the overall 2SLAQ LRG sample have recently been studied by 
Ross et al.\ (2006), who note that the 2SLAQ spectroscopic 
sample is biased against close ($<$30\,arcsec) pairs of galaxies because of the minimum 
fibre separation set by the 2dF positioner (the so-called ``fibre collision problem'').  
In general, the 2SLAQ spectroscopic catalogue omits about 65\% of objects which are 
within 30\,arcsec of another 2SLAQ galaxy, about half of those within 1\,arcmin and 
virtually none of those with separations larger than about 3\,arcmin. 
Figure \ref{fig.cl} shows an example of a pair of 2SLAQ radio 
galaxies which are almost certainly members of a cluster at $z\sim0.6$.  In this 
case the angular separation is just over 30\,arcsec, and both galaxies were observed 
spectroscopically.  The velocity difference between these two radio galaxies 
is 420\,km\,s$^{-1}$, typical of the velocity dispersion of galaxies 
in a modest-sized cluster or large galaxy group.   

\subsection{The two--point correlation function for 2SLAQ radio galaxies} 

\begin{figure}
\vspace*{8.5cm}
\includegraphics{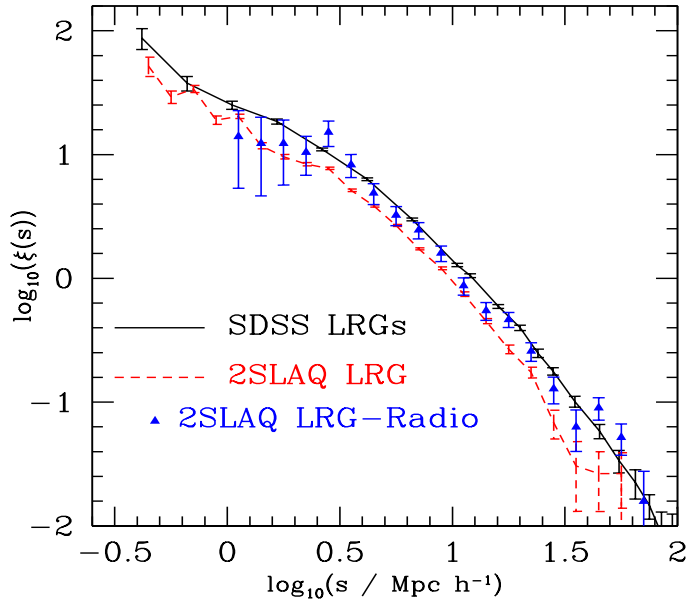}
\caption{The redshift--space correlation function $\xi$(s), between the radio--detected 2SLAQ LRGs 
and the full LRg sample (filled triangles). The function has been corrected for 2dF 
``fibre collision'' effects on small scales, as described by Ross et al.\ 2006).  For comparison, 
the solid line shows the redshift--space correlation function for SDSS LRGs (Zehavi et al.\ 2005; 
Eisenstein et al.\ 2005) and the dashed line the finction for all 2SLAQ LRGs (Ross et al.\ 2006). 
  }
\label{fig.xis} 
\end{figure} 

We investigated the clustering properties of the 2SLAQ radio galaxies in a more quantitative 
way by calculating the redshift--space two--point correlation function between the 2SLAQ 
radio galaxies and the full LRG sample.  The triangular points 
in Figure \ref{fig.xis} show the measured redshift--space correlation $\xi$(s) between 
303 Sample 8 radio galaxies from Table 3 and a set of 8656 LRGs which are also in Sample 8. 
Further details of the calculations and sample definition are given by Ross et al.\ (2006). 
The dashed line in Figure \ref{fig.xis} is the 2SLAQ LRG autocorrelation measured by Ross et 
al.\ (2006), and the solid line shows the result for the lower--redshift SDSS LRGs 
from Zehavi et al.\ (2005) on intermediate-scales and Eisenstein et al.\ (2005) at the
large-scales. 

As discussed by Ross et al.\ (2006), the 2SLAQ LRG measurements lie below the SDSS LRG line. 
This should not necessarily be taken as evidence of evolution in the clustering properties 
of LRGs.  Although the SDSS LRG survey is at lower redshift than 2SLAQ, it has stricter colour 
selection criteria and so excludes many of the bluer, less luminous LRGs 
which are present in the 2SLAQ sample.  
The SDSS LRGs may therefore appear more clustered simply because they are on average 
more luminous (and hence more biased) than the 2SLAQ LRGs.  

Interestingly, the 2SLAQ radio galaxies have a clustering strength which is higher 
than the 2SLAQ LRG population as a whole, but lower than the SDSS LRGs. 
There is no $a\ priori$\ reason to expect that radio galaxies will be clustered 
differently from other luminous early--type galaxies.  
Ledlow \& Owen (1995) showed that the probability of a luminous early--type galaxy 
hosting a radio source is unaffected by its clustering environment, and Blake \& Wall 
(2002) found that the angular clustering of NVSS radio sources is similar to that of 
the general galaxy population.  We have shown in Table \ref{tab.magstats} that the 
2SLAQ radio galaxies are significantly more luminous than the 2SLAQ LRG population 
as a whole. 
It therefore seems likely that the higher clustering 
strength of the 2SLAQ radio galaxies can be explained simply by their higher luminosity, 
though it would be interesting to test this more explicitly. 

\subsection{Low--power radio galaxies and the FR\,I/II break} 
\begin{figure}

\vspace*{8.5cm}
\includegraphics{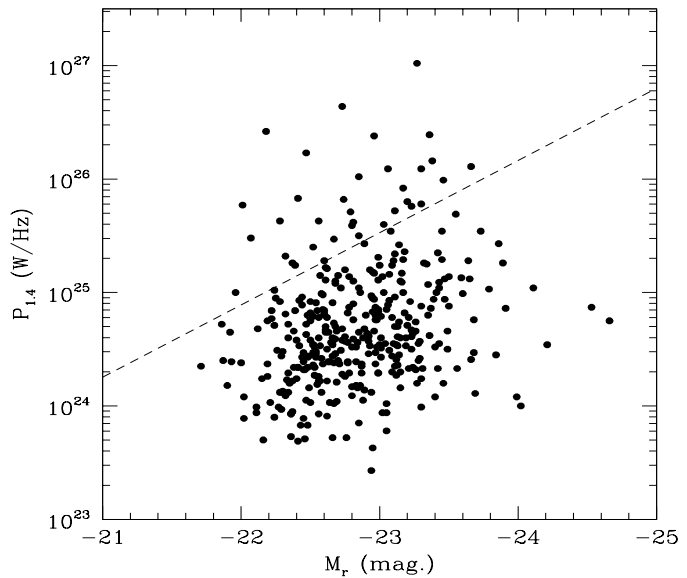}
\caption{Comparison of 1.4\,GHz radio luminosity and $r$--band absolute 
magnitude for 2SLAQ radio galaxies.  The dashed line represents the division 
between local FR\,I and FR\.II radio galaxies found by Ledlow \& Owen (1996), with 
FR\,II radio galaxies lying above the line and FR\,Is below.   }
\label{fig.fr2} 
\end{figure}

The lowest--power radio galaxies in our 2SLAQ sample 
have radio luminosities similar to those of FR\,I radio galaxies in the local universe 
(see Figure \ref{fig.fr2}). The currently--available FIRST and 
NVSS radio images do not have high-enough resolution and sensitivity to allow one  
to distinguish reliably between FR\,I and FR\,II morphologies for most radio 
galaxies in the 2SLAQ sample.  At this stage, therefore, it is unclear whether the 
different rates of cosmic evolution which we find for low--power and high--power 
radio galaxies represent differences between the FR\,I and FR\,II radio--galaxy 
populations, or simply a general tendency for more luminous radio galaxies to 
undergo more rapid luminosity evolution. 
Deeper high--resolution radio observations 
of the galaxies in Table 3 are needed to address this question further. 

\subsection{AGN heating and the evolution of massive galaxies}
Feedback from AGN heating, and specifically mechanical heating of the interstellar gas 
by radio jets, is a recent addition to hierarchical models for the formation of massive 
galaxies (e.g. Croton et al.\ 2006; Bower et al.\ 2006).  In these models, re-heating of 
cooling gas by an AGN prevents low--redshift star formation from occurring in massive 
early--type galaxies, making it possible for hierarchical models to reproduce the 
observed colours and luminosity function of early--type galaxies. 

Best et al.\ (2006) recently derived an empirical conversion between the observed radio 
luminosity of an AGN and the mechanical energy input into the parent galaxy, and found 
that the total energy input scaled roughly as P$^{0.4\pm0.1}$, where P is the total 
radio luminosity.  If this is correct, then the 
global rate of AGN heating in the local universe is dominated by the contribution of 
low--luminosity ($10^{22}<$P$_{1.4}<10^{24}$\,W\,Hz$^{-1}$) radio galaxies, since these   
make up the great majority of radio-loud AGN. 

Our results imply that the low--power radio galaxies associated with LRGs in the local 
universe would typically have been at least twice as powerful at z$\sim$0.5 and 
(by extrapolation) perhaps five times as powerful at $z\sim1$.  If the same kind of 
evolution also occurs for radio galaxies weaker than those observed here, then the 
total energy input from AGN heating in massive early--type galaxies would have been 
up to 50\% higher at $z\sim0.5$ than in the local universe, and perhaps twice as high 
at $\sim1$.  

\begin{figure}
\vspace*{8.5cm}
\includegraphics{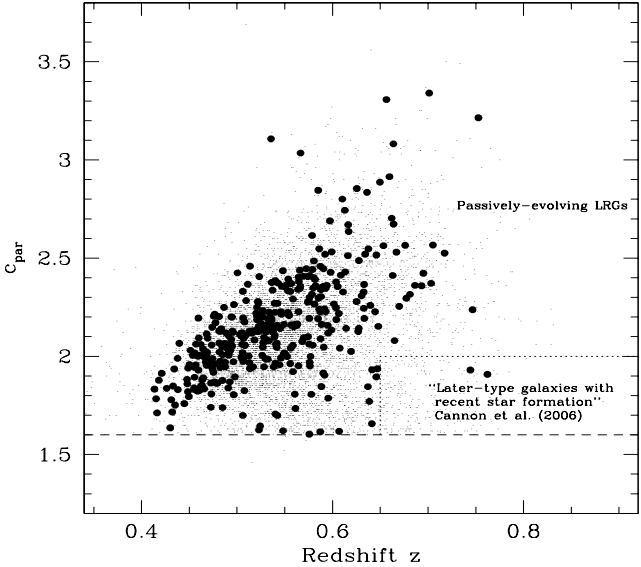}
\caption{Values of the photometric parameter $c_{\rm par}=0.7(g-r)+1.2(r-i-0.18)$ 
plotted against redshift for the full 2SLAQ LRG sample (small dots) and the 2SLAQ 
radio galaxies (large filled circles). 
The horizontal line at  $c_{\rm par}=1.6$ marks the expected division between 
passive (red) and star--forming (blue) galaxies in this redshift range, and 
the box marked by dotted lines at $z>0.65$ and $c_{\rm par}<2.0$ is the 
region identified by Cannon et al. as containing a population of later--type 
high--redshift galaxies with recent star formation.
 }
\label{fig.cpar}
\end{figure} 

\subsection{Effects of the 2SLAQ colour and luminosity cuts} 
The rate of cosmic evolution measured for low--power radio galaxies in this study 
is a lower limit, since the 2SLAQ LRG sample has a strict colour cutoff 
(as plotted in Figure \ref{fig.phot}) whereas no such colour restriction 
was applied to the local (6dFGS) radio galaxies.  If there were an additional 
population of blue radio galaxies at redshift $z\sim0.55$, they might be 
removed by the 2SLAQ colour selection and so missed from our analysis.  
Figure \ref{fig.cpar} plots the parameter $c_{\rm par}=0.7(g-r)+1.2(r-i-0.18)$ 
against redshift for our radio--galaxy sample as well as the full 2SLAQ LRG 
sample.  Any recent or on-going star formation would scatter LRGs 
to lower values of $c_{\rm par}$ in this plot, as discussed by Cannon et 
al.\ (2006) and Roseboom et al.\ (2006).  Well over 90\% of 2SLAQ 
radio galaxies lie close to the locus of passively--evolving galaxies seen 
in the main 2SLAQ LRG sample, suggesting that they have undergone 
little or no recent star formation. 


Cannon et al.\ (2006) have pointed out that the highest--redshift ($z>0.65$)  
2SLAQ LRGs have a bimodal distribution in $c_{\rm par}$, and attribute this 
to later--type luminous galaxies with recent star formation which have spilled 
across the colour--selection boundary.  The region occupied by these objects 
is shown in Figure \ref{fig.cpar}.  Radio galaxies with ongoing star formation 
should also be easy to identify in this region, but very few 
are seen (fewer than 5\% of 2SLAQ radio galaxies lie near the $c_{\rm par}$ 
boundary in Figure \ref{fig.cpar}).  We therefore estimate that only a small 
number of star--forming radio galaxies are being excluded from our sample 
by the 2SLAQ colour cuts. 

A second consideration is the cutoff in optical luminosity set by the 2SLAQ 
$i=19.8$\,magnitude limit for 2dF spectroscopy. The effect of this 
is to set a fairly sharp boundary in at M$_r\simeq-22.0$, as can be  
seen in Figure \ref{fig.fr2}.  This corresponds to a lower limit of about 
$2\times10^{11}$\,M$_\odot$ in stellar mass if we adopt the values of 
log\,(M/L)$_r\sim0.6$  used by Kauffmann et al.\ (2003).  

In the local universe, as discussed in \S5.3.1, this cutoff would have 
little or no effect on measurements of the radio luminosity function. 
This is confirmed by the recent work of Best et al.\ (2005b) who find that 
fewer than 1\% of SDSS galaxies with stellar masses below $2\times10^{11}$\,M$_\odot$ 
host a radio--loud AGN, even at radio luminosities an order of magnitude 
lower than those considered here.  At $z\sim0.55$, however, this is 
not necessarily the case, and it is possible that galaxies below the 2SLAQ 
cutoff in optical luminosity could host significant numbers of radio galaxies. 
Because of this caveat, we stress that the radio luminosity functions derived 
in this paper apply to optically bright early--type galaxies.  If less--luminous 
galaxies also contribute significantly to the radio luminosity function in the 
2SLAQ redshift range, then the overall rate of cosmic evolution in the 
radio--galaxy population would be even higher than the values derived in \S5.3.

\section{Summary and future work}
The main results of this study are: 
\begin{itemize}
\item 
We have measured an accurate radio luminosity function for luminous early--type 
galaxies in the redshift range $0.4<z<0.7$, and compared this with the radio 
luminosity function for the same population in the local 
universe. This allows us to measure the cosmic evolution of AGN-related radio 
emission in a single, well-defined galaxy population. 
\item
We find clear evidence that low--power radio galaxies (i.e. those in the 
luminosity range covered by local FR\,I radio galaxies) undergo significant 
cosmic evolution over the redshift range $0<z<0.7$.  This evolution is 
well--fitted by pure luminosity evolution of the form (1+$z$)$^k$ (where 
$k$=2.0$\pm$0.3) for galaxies with radio luminosities between 10$^{24}$ 
and 10$^{26}$ W\,Hz$^{-1}$. The overall rate of evolution may be higher than 
this if some low--power radio galaxies at $z\sim0.55$ have been excluded  
by the $i=19.8$\,mag. cutoff of the 2SLAQ spectroscopic sample. 
\item 
The most powerful radio galaxies in our sample 
(those with 1.4\,GHz radio powers above 10$^{26}$\,W\,Hz$^{-1}$) may evolve more 
rapidly than lower--luminosity radio galaxies over the redshift range $0<z<0.7$.
\item
The radio detection rate is highest for the most luminous 2SLAQ LRGs, 
and the median absolute magnitude of our radio--galaxy sample is 0.4\,mag 
brighter than that of the 2SLAQ LRG population as a whole. 
\item 
The 2SLAQ radio galaxies are more strongly clustered than the overall 2SLAQ LRG sample, 
possibly because they are on average more luminous.  
\item 
Most 2SLAQ radio galaxies show no obvious emission lines in their optical spectra.  
The fraction showing 3727\AA\ [O\,II] emission is slightly higher than in the 
2SLAQ LRG sample as a whole, this but this may be due to the higher optical luminosity 
of the radio--galaxy subsample and a more detailed analysis is needed. 
About 3\% of the 2SLAQ radio galaxies show high--ionization emission lines of 
3426\AA\ [Ne\,V] and/or 3870\AA\ [Ne\,III].  
\end{itemize}

This paper presents a first analysis of the evolving radio--galaxy population 
out to $z\sim0.7$.  A second paper 
(Johnston et al.\ 2007, in preparation) will study the stellar populations of radio--loud 
LRGs in the 2SLAQ survey.  Further radio observations of the 2SLAQ area would be valuable, 
both at higher spatial resolution (to test whether the FR\,I/FR\,II divide seen in 
Figure \ref{fig.fr2} evolves with redshift) and to fainter flux limits (to probe further 
down the radio luminosity function).  Since the 2SLAQ survey also targeted QSOs over 
a redshift range which overlaps that covered by the LRG survey, a comparison of 
the relative numbers of radio galaxies and radio--loud QSOs in the same redshift 
range would give a useful test of the unified model for radio--loud AGN. 

The new 2dF AAOmega spectrograph recently commissioned at the Anglo--Australian Telescope 
has significantly higher efficiency than the system used for the 2SLAQ survey, and now 
makes it possible to extend the 2SLAQ LRG survey to higher redshifts and/or larger samples.  
In particular, a large spectroscopic redshift survey of LRGs out to $z\sim 1$ appears to 
be feasible in the near future. 

New radio surveys will be needed to identify and study low--power 
radio galaxies beyond $z\sim0.7$ in conjunction with the new generation 
of large-area spectroscopic surveys, though the existing FIRST and NVSS surveys will 
still be able to detect many of the most powerful radio galaxies out to $z\sim1$. 
A new radio survey would ideally be able to detect 
sources well below 1\,mJy over a wide area of sky (at least 50 square degrees), 
and achieve both good positional accuracy and good surface--brightness sensitivity. 
While this is time-consuming at present, it should be easier when new wide--band
correlators come into use at both the Australia Telescope Compact Array (ATCA) and the 
VLA over the next few years. 

\section{Acknowledgements}
We thank the staff of the Anglo--Australian Observatory for their help in maintaining 
and running the 2dF spectrograph throughout the course of the 2SLAQ survey. 
%
EMS acknowledges support from the Australian Research Council through the award of an ARC 
Australian Professorial Fellowship. 

Funding for the SDSS and SDSS-II has been provided by the Alfred P. Sloan Foundation, the
Participating Institutions, the National Science Foundation, the U.S. Department of Energy, 
the National Aeronautics and Space Administration, the Japanese Monbukagakusho, the Max Planck 
Society, and the Higher Education Funding Council for England. The SDSS Web Site is http://www.sdss.org/.

The SDSS is managed by the Astrophysical Research Consortium for the Participating Institutions. 
The Participating Institutions are the American Museum of Natural History, Astrophysical Institute 
Potsdam, University of Basel, Cambridge University, Case Western Reserve University, University of Chicago, 
Drexel University, Fermilab, the Institute for Advanced Study, the Japan Participation Group, 
Johns Hopkins University, the Joint Institute for Nuclear Astrophysics, the Kavli Institute for 
Particle Astrophysics and Cosmology, the Korean Scientist Group, the Chinese Academy of Sciences 
(LAMOST), Los Alamos National Laboratory, the Max-Planck-Institute for Astronomy (MPIA), 
the Max-Planck-Institute for Astrophysics (MPA), New Mexico State University, Ohio State University, 
University of Pittsburgh, University of Portsmouth, Princeton University, the United States Naval 
Observatory, and the University of Washington.

We thank the referee for several useful comments which improved the final version of this paper.


\begin{thebibliography}{99}  
\bibitem{} Auriemma, C., Perola, G.C., Ekers, R.D., Fanti, R., Lari, C., Jaffe, W.J., 
 Ulrich, M.H., 1977, A\&A, 57, 41 
\bibitem{} Becker, R.H., White, R.L., Helfand, D.J., 1995, ApJ, 450, 559 
\bibitem{} Best, P.N., Kauffmann, G., Heckman, T, Ivezic, Z., 2005a, MNRAS 362, 9
\bibitem{} Best, P.N., Kauffmann, G., Heckman, T, Brinchmann, J., Charlot, S., Ivezic, Z., 
   White, S.D.M., 2005b, MNRAS 362, 25 
\bibitem{} Best, P.N., Kaiser, C.R., Heckman, T.M., Kauffmann, G., 2006, MNRAS, 368, L67 
\bibitem{} Bicknell, G.V., 1995, ApJS, 101, 29 
\bibitem{} Binney, J., Tabor, G., 1995, MNRAS 276, 663 
\bibitem{} Birzan, L., Rafferty, D.A., McNamara, B.R., Wise, M.W., Nulsen, P.E.J., 2004, ApJ 607, 800 
\bibitem{} Blake, C., Wall, J., 2002, MNRAS, 337, 993 
\bibitem{} Blanton, E.L., Gregg, M.D., Helfand, D.J., Becker, R.H., White, R.L., 2003, AJ, 125, 1635 
\bibitem{} Blundell, K.M., Rawlings, S., Willott, C.J., Kassim, N.E., 
  Perley, R., 2002, New Astronomy Reviews, 46, 75  
\bibitem{} Bower, R.G., Benson, A.J., Malbon, R., Helly, J.C., Frenk, C.S., Baugh, C.M., 
Cole, S., Lacey, C.G., 2006, MNRAS, in press
\bibitem{} Boyle, B.J., Shanks, T., Peterson, B.A., 1988, MNRAS, 238, 957 
\bibitem{} Bressan, A., Chiosi, C., Fagotto, F., 1994, ApJS, 94, 63 
\bibitem{} Brown, M.J.I., Webster, R.L., Boyle, B.J., 2001, AJ, 121, 2381 
\bibitem{} Cannon, R.D. et al., 2006, MNRAS, 372, 425 
\bibitem{} Clewley, L., Jarvis, M.J., 2004, MNRAS, 352, 909 
\bibitem{} Colless, M. et al., 2001, MNRAS, 328, 1039 
\bibitem{} Condon, J.J., Cotton, W.D., Greisen, E.W., Yin, Q.F., Perley, R.A., Taylor, G.B., 
   Broderick, J.J., 1998, AJ, 115, 1693 
\bibitem{} Condon, J.J., 1992, ARA\&A 30, 575
\bibitem{} Croton, D.\ et al., 2006, MNRAS, 365, 11
\bibitem{} Doroshkevich, A.G., Longair, M.S., Zeldovich, Y.B., 1970, MNRAS 147, 139 
\bibitem{} Dunlop, J.S., Peacock, J.A., 1990, MNRAS 247, 19 
\bibitem{} Eisenstein, D.J. et al., 2001, AJ, 122, 2267 
\bibitem{} Eisenstein, D.J. et al., 2005, ApJ, 633, 560
\bibitem{} Fanaroff, B,L, Riley, J.M., 1974, MNRAS, 167, 31 
\bibitem{} Fukugita, M., Ichikawa, T., Gunn, J.E., Doi, M., Shimasaku, K., Schneider, D.P., 
   1996, AJ, 111, 1748
\bibitem{} Hopkins, A.M., Afonso, J., Chan, B., Cram, L.E., Georgakakis, A., Mobasher, B., 2003, AJ, 125, 465 
\bibitem{} Hopkins, A.M., 2004,ApJ 615, 209  
\bibitem{} Ivezic, Z. et al., 2002, AJ, 124, 2364 
\bibitem{} Jackson, C.A., Wall, J.V., 1999, MNRAS, 304, 160 
\bibitem{} Jones, D.H. et al., 2004, MNRAS, 355, 747 
\bibitem{} Jones, D.H., Peterson, B.A., Colless, M., Saunders, W., 2006, MNRAS, 369, 25 
\bibitem{} Kauffmann, G. et al., 2003, MNRAS, 341, 33
\bibitem{} Laing, R.A., Riley, J.M., Longair, M.S., 1983, MNRAS 204, 1511 
\bibitem{} Lampton, M., Margon, B., Bowyer, S., 1976, ApJ, 208, 177 
\bibitem{} Large, M.I., Mills, B.Y., Little, A.G., Crawford, D.F., Sutton, J.M., 
  1981, MNRAS, 194, 693 
\bibitem{} Ledlow, M.J., Owen, F.R., 1995, AJ, 109, 853 
\bibitem{} Ledlow, M.J., Owen, F.R., 1996, AJ, 112, 9
\bibitem{} Lewis, I.J. et al., 2002, MNRAS, 333, 279 
\bibitem{} Longair, M.S., 1966, MNRAS 133, 421 
\bibitem{} Magorrian, J. et al., 1998, AJ, 115, 2285 
\bibitem{} Mauch, T., Sadler, E.M., 2006, MNRAS, submitted 
\bibitem{} Owen, F.R., Ledlow, M.J., Keel, W.C., 1995, AJ, 109, 14  
\bibitem{} Peacock, J.A., 1999, `Cosmological Physics', Cambridge University Press, p.\ 443--445
\bibitem{} Phillips, M.M., Jenkins, C.R., Dopita, M.A., Sadler, E.M., Binette, L., 1986, AJ, 91, 1062
\bibitem{} Rawlings, S., Jarvis, M.J., 2004, MNRAS 355, L9
\bibitem{} Rocca--Volmerange, B., Le Borgne, D., De Breuck, C., Fioc, M., Moy, E., 2004, A\&A, 415, 931 
\bibitem{} Roseboom, I.G. et al., 2006, MNRAS, in press 
\bibitem{} Ross, N.P. et al., 2006, MNRAS, submitted 
\bibitem{} Sadler, E.M., Jenkins, C.R., Kotanyi, C.G., 1989, MNRAS, 240, 591 
\bibitem{} Sadler, E.M. et al., 2002, MNRAS, 329, 227 
\bibitem{} Schmidt, M., 1968, ApJ, 151, 393 
\bibitem{} Snellen, I.A.G., Best, P.N., 2001, MNRAS, 328, 897 
\bibitem{} Springel, V., Di Matteo, T., Hernquist, L., 2005, MNRAS 361, 776 
\bibitem{} Stoughton, C.\ et al., 2002, AJ, 123, 485 
\bibitem{} Veilleux, S., Osterbrock, D.E., 1987, ApJS, 63, 295 
\bibitem{} Wake, D.A. et al., 2006, MNRAS, 372, 537 
\bibitem{} Wall, J.V., 1990, Phil.\ Trans.\ R.\ Soc., A296, 367 
\bibitem{} Willott, C.J., Rawlings, S., Blundell, K.M., Lacy, M., Eales, S.A., 2001, MNRAS, 322, 536
\bibitem{} York, D.G. et al., 2000, AJ 120, 1579
\bibitem{} Zehavi, I. et al., 2005, ApJ, 608, 16

\end{thebibliography}
\end{document}